\documentclass[a4paper,11pt]{article}
\pdfoutput=1
\usepackage{jcappub} % for details on the use of the package, please see the JINST-author-manual
\usepackage{lineno}
\usepackage{aas_macros}
\usepackage[normalem]{ulem}
\usepackage{comment}

\usepackage[T1]{fontenc}
\usepackage{array}
\usepackage{makecell}

\usepackage{amsmath}
\usepackage{amssymb}
\usepackage{hyperref}

\arxivnumber{2402.12452}
\title{\boldmath Numerical Challenges in Modeling Gravothermal Collapse in Self-Interacting Dark Matter Halos}

\author[a]{Igor Palubski,}
\author[b,d,e]{Oren Slone,}
\author[a]{Manoj Kaplinghat,}
\author[b,c]{Mariangela Lisanti,}
\author[f]{Fangzhou Jiang}

\affiliation[a]{Department of Physics and Astronomy, University of California, Irvine, CA 92617, USA}
\affiliation[b]{Department of Physics, Princeton University, Princeton, NJ 08544, USA}
\affiliation[c]{Center for Computational Astrophysics, Flatiron Institute, New York, NY 10010, USA}
\affiliation[d]{Center for Cosmology and Particle Physics, Department of Physics, New York University, New York, NY 10003, USA}
\affiliation[e]{C.~N.~Yang Institute for Theoretical Physics, Stony Brook University, Stony Brook, NY 11794, USA}
\affiliation[f]{Kavli Institute for Astronomy and Astrophysics, Peking University, Beijing 100871, China}

\emailAdd{ipalubsk@uci.edu}

\abstract{
When dark matter has a large cross section for self scattering, halos can undergo a process known as gravothermal core collapse, where the inner core rapidly increases in density and temperature.  
To date, several methods have been used to implement Self-Interacting Dark Matter~(SIDM) in N-body codes, but there has been no systematic study of these different methods or their accuracy in the core-collapse phase. In this paper, we compare three different numerical implementations of SIDM, including the standard methods from the GIZMO and Arepo codes, by simulating idealized dwarf halos undergoing significant dark matter self interactions ($\sigma/m = 50$~cm$^2$/g). 
When simulating these halos, we also vary the mass  
resolution, time-stepping criteria, and gravitational force-softening scheme. The various SIDM methods lead to distinct differences in a halo's evolution during the core-collapse phase, as each results in spurious scattering rate differences and energy gains/losses. 
The use of adaptive force softening for gravity can lead to numerical heating that artificially accelerates core collapse, while an insufficiently small simulation time step can cause core evolution to stall or completely reverse. Additionally, particle numbers must be large enough to ensure that the simulated halos are not sensitive to noise in the initial conditions. Even for the highest-resolution simulations tested in this study ($10^6$ particles per halo), we find that variations of order $10\%$ in collapse time are still present. 
The results of this work underscore the sensitivity of SIDM modeling on the choice of numerical implementation and motivate a careful study of how these results generalize to halos in a  cosmological context.
}

\begin{document}
\maketitle
\flushbottom

\section{Introduction}

Dark matter~(DM) self interactions provide the means to transfer heat and mass across a halo~\citep{Spergel2000}. When the self interactions are strong enough, they can impact the structure, morphology and diversity of galaxies and their satellites---see Refs.~\cite{Tulin:2017ara,Adhikari:2022sbh} for reviews. This provides an exciting opportunity for Self-Interacting Dark Matter~(SIDM) to be distinguished from collision-less Cold Dark Matter~(CDM) through purely gravitational interactions on galactic and sub-galactic scales. 
Robust interpretations of such observations necessarily rely on careful modeling of galaxy evolution in these different DM frameworks.  Towards this goal, this paper provides a first study of numerical challenges associated with simulating SIDM halos whose cores contract to high densities.

The heat transfer provided by collisions of DM particles allows the inner-most regions of a halo to heat up. During this core-expansion phase, the DM particles at the halo's center acquire kinetic energy and expand their orbits. This reduces the core density for as long as heat is transferred inward~\citep{Kochanek2000,Yoshida2000,Koda2011,Vogelsberger2012,Rocha2013}. Eventually, the core becomes hotter than the outer regions, and the heat transfer flips so that heat is transferred from the isothermal core to the outer halo. This leads to the core shrinking in spatial size and increasing in density~\citep{BalbShap2002,Elbert:2014bma}---a slow process because the temperature gradient in the inner halo is extremely shallow. The net effect of the shrinking core and outward heat flow is an increase in the temperature of the core as long as the system is in hydrostatic equilibrium. The increase in core temperature further facilitates the outward heat flow resulting in the runaway process called gravothermal core collapse. This process was first studied in the context of globular clusters---see e.g., Ref.~\cite{Lyndel1980}---and was later applied to the study of SIDM~\citep{BalbShap2002,Koda2011,Essig2019,Nishikawa2020,Yang:2022hkm,Yang2023,Yang:2023jwn,Jiang2023,Zhong:2023yzk}.

Self-scattering interactions with a cross section as small as $3~\text{cm}^2/$g have been shown to increase the diversity of rotation curves in the inner parts and bring them closer in agreement with observations~\citep{Ren:2018jpt,Zentner:2022xux}. 
However, recent studies show that an interaction cross section larger than $10~\text{cm}^2/$g~\citep{Kaplinghat2019,Zavala:2019sjk,Kahlhoefer2019,Nishikawa2020} and as high as $100~\text{cm}^2/$g~\citep{Correa2021,Turner2021,Yang2022} at the dwarf spheroidal galaxy velocity scales of 10--30~km/s could reproduce the diversity of central densities measured in the satellite galaxies of the Milky Way. SIDM models that do not allow for core collapse are disfavored by the Milky Way satellite kinematic data~\cite{Slone:2021nqd,Silverman2022}. Core collapse in satellites could contribute to the large galaxy-galaxy strong lensing excess observed in several galaxy clusters~\citep{Meneghetti2022,Yang2021} and the anomalous densities of subhalos detected in galaxy-galaxy lensing~\citep{Minor2021,Sengul:2022edu,Zhang:2023wda}. Core collapse could additionally provide the seeds for Super Massive Black Holes~(SMBH) observed at high redshift $(z \geq 7)$~\citep{Pollack2015,Damico2018,Latif2019,Choquette2019,Feng2021}. Such early core collapse would require a large interaction cross section, or dissipative scattering, or a more complex dark sector, e.g mirror DM~\citep{Blinnikov1983,Klhopov1991,Foot2004,Berezhiani2005,Pollack2015,Choquette2019}.

Modeling galaxy formation in the context of SIDM requires supplementing gravitational scattering with DM self-scattering in standard N-body codes~\citep{Kochanek2000,Yoshida2000,Koda2011,Vogelsberger2012,Rocha2013,Fry2015,Robles2017,Robertson2017a,Robertson2017b,Nadler2020,Banerjee2020,Correa2022}. Three commonly used implementations include the Spline method written into Arepo~\citep{Vogelsberger2012}, the Kernel-Overlap method in GIZMO~\citep{Rocha2013,Robles2017}, and the Top-Hat method~\cite{Kochanek2000,Yoshida2000,Robertson2017a,Robertson2017b}. These methods differ, in part, in how they account for nearest neighbors and how they implement force softening. The authors of these codes tested the methods by verifying the scattering rates in DM halos with a Hernquist~\citep{Hernquist1990} or Navarro-Frenk-White~(NFW)~\citep{Navarro1997} profile, or by shooting individual DM particles at a uniform field of background particles. 
Ref.~\cite{Meskhidze2022} performed the first comparison of the N-body codes Arepo and GIZMO in their base configuration for an isolated SIDM halo in the core-expansion phase. In their high-resolution simulations, employing fixed force softening much smaller than the DM core size, they found significant differences in the overall scattering rate between the two methods and noted up to 30\% differences in their density profiles in the core-expansion phase. However, no convergence study has yet been performed that compares different SIDM implementations in the same N-body code and also pushes this comparison deep into the core-collapse phase.

In this paper, we perform a detailed comparison of the three aforementioned SIDM implementations and discuss the appropriate choices of numerical parameters and mass-resolution limitations. In particular, we consider how choices in number of particles, time stepping, and gravitational force-softening schemes impact a halo's core density and collapse time scales.  As a specific example, we focus on an isolated dwarf galaxy of mass $1.15 \times 10^9 M_{\odot}$ at both a high and low concentration.  These two cases cover both early and late core collapse. For a cross section of $50\;\rm cm^2/g$, we investigate how different SIDM implementations reproduce the core collapse of a DM halo by comparing the evolution of core density and velocity dispersion. As will be demonstrated, the evolution is highly sensitive to the numerical implementation of both the gravitational and self-scattering processes.

This paper is organized as follows. The details of the simulation implementation are described in Sec.~\ref{sec:framework}. Section~\ref{sec:numerical} discuses how halo evolution is affected by the SIDM implementation, mass resolution (number of particles), time-stepping criteria and gravitational force-softening scheme.  Section~\ref{fluid} comments on how our numerical results compare with a fluid description of core-collapsing halos. We conclude in Sec.~\ref{sec:conclusions}.  An appendix is included with some supplementary figures.

\section{Simulation Framework}
\label{sec:framework}

The DM-only halos studied in this work are evolved using the gravity-tree solver in GIZMO~\citep{Hopkins2015}. 
As a concrete case study, we consider a low- and high-concentration variant of a $M_{\rm 200} = 1.15 \times 10^9 M_{\odot}$ isolated halo with virial radius $R_{\rm 200} = 22.1$~kpc. The initial conditions for these halos are set using \texttt{Spheric}~\citep{Garrison2013} and assume an NFW density profile with some scale radius, $r_{\rm s}$, and density, $\rho_{\rm s}$. All  halos are exponentially truncated at $r_{\rm trunc} = 23.6$~kpc beyond the virial radius (see Ref.~\cite{Zemp2008} for details on the truncation form), similar to previous studies~\citep{Zeng2022}.\footnote{Ref.~\cite{Nishikawa2020} found that the truncation radius has no effect on a halo's evolution so long as it is larger than several times the scale radius.} The high-concentration halo has $\rho_{\rm s} = 1.04 \times 10^8 M_{\odot}/\rm kpc^{3}$ and $r_{\rm s} = 0.715$~kpc, which corresponds to a concentration of  $c_{\scriptstyle{200}} = 31$. The low-concentration halo has parameters of $\rho_{\rm s} = 2.73 \times 10^7 M_{\odot}/\rm kpc^{3}$ and $r_{\rm s} = 1.18$~kpc, corresponding to $c_{\scriptstyle{200}} = 19$. \textbf{This study takes an SIDM cross section per unit DM particle mass of $\sigma/m = 50$~cm$^2$/g as a benchmark scenario.} These parameters result in a core-collapse time of $
\sim 3.5$~($14$)~Gyr for the high~(low) concentration case, providing a good example of early versus late collapse in cosmological time. 

For the two halos under consideration, we vary several different inputs to the numerical modeling, including the DM scattering implementation~(Sec.~\ref{sec:implementation}), force-softening scheme~(Sec.~\ref{sec:forcesoft}), and time-stepping criteria (Sec.~\ref{sec:timestep_implementation}).  Table~\ref{tab:sims} summarizes the simulations used in this work.  The halos are run at low, medium, and high resolution, which correspond to $N_{\rm p} =  3\times10^4$, $5\times10^5$, and $10^6$ total gravitationally-bound particles in the halo, respectively. Additionally, all simulations are run with the same number of cores to minimize the influence of hardware on the results.

\begin{table*}
\renewcommand{\arraystretch}{1.5}
\centering
\begin{tabular}{c c c c c c c c}
\Xhline{2\arrayrulewidth}
$c_{\scriptstyle{200}} $ & $r_{\rm s}$ & $\rho_{\rm s}$ & $N_p$ & $h_{g,i}$ & $h_{s,i}$ & $\kappa$ & $\eta$\\
 & $\left[ \text{kpc} \right]$ & $\left[M_\odot / \rm kpc^3\right]$ & & $\left[ \text{kpc} \right]$ & $\left[ \text{kpc} \right]$&  &\\
\Xhline{2\arrayrulewidth}
19 (low) & $1.18$ & $2.73 \times 10^{7}$ & $3 \times 10^{4}$& adaptive & adaptive &$0.02 \, \& \, 0.002$ & $0.02$ \\
19 (low) & $1.18$ & $2.73 \times 10^{7}$ & $5 \times 10^{5}$& adaptive & adaptive & $0.02 \, \& \, 0.002$ & $0.02$ \\
19 (low) & $1.18$ & $2.73 \times 10^{7}$ & $1 \times 10^{6}$& adaptive & adaptive & $0.02 \, \& \, 0.002$ & $0.02$ \\
\hline
19 (low) & $1.18$ & $2.73 \times 10^{7}$ & $3 \times 10^{4}$& adaptive & adaptive &   $0.002$ & $0.002$ \\
19 (low) & $1.18$ & $2.73 \times 10^{7}$ & $5 \times 10^{5}$& 0.0353 & adaptive &  $0.002$ & $0.002$ \\
19 (low) & $1.18$ & $2.73 \times 10^{7}$ & $1 \times 10^{6}$& 0.0353 & adaptive & $0.002$ & $0.02$ \\
\hline
31 (high) & $0.715$ & $1.04 \times 10^{8}$ & $3 \times 10^{4}$& adaptive & adaptive & $0.002$ & $0.02$\\
31 (high) & $0.715$ & $1.04 \times 10^{8}$ & $5 \times 10^{5}$& adaptive & adaptive & $0.002$ & $0.02$\\
31 (high) & $0.715$ & $1.04 \times 10^{8}$ & $1 \times 10^{6}$& adaptive & adaptive & $0.002$ & $0.02$\\
\Xhline{2\arrayrulewidth}
\end{tabular}
\caption{Halo and simulation parameters used in this work, including the concentration~($c_{\scriptstyle{200}} = R_{200}/r_{\rm s}$), NFW scale radius~($r_{\rm s}$) and density~($\rho_{\rm s}$), number of particles~($N_p$), force-softening length for gravitational~($h_{g,i}$) and self~($h_{s,i}$) interactions, time-stepping criterion~($\kappa$), and  tolerance parameter~($\eta$).  The parameter $\kappa$ corresponds to the maximum probability that two particles scatter with each other in a given time step and is defined in Sec.~\ref{sec:timestep_implementation}. The parameter $\eta$ captures  the fraction of the gravitational softening length that a particle can travel in a specified time step; it is discussed in Sec.~\ref{sec:forcesoft}.  All of the listed configurations are simulated using the three SIDM methods described in Sec.~\ref{sec:implementation}: Kernel Overlap, Spline, and Top Hat.}
\label{tab:sims}
\end{table*}

\subsection{Implementation of Dark Matter Scattering}
\label{sec:implementation}

The interactions are implemented using three different SIDM methods from  the literature: the Kernel-Overlap, Spline, and Top-Hat methods. The Kernel-Overlap method is available in the public version of GIZMO, while the Top-Hat and Spline methods have been re-implemented for this work.

\subsubsection{Kernel-Overlap Method}
\label{sec:kerneloverlap}

The Kernel-Overlap procedure applies a scattering method derived from the collisional Boltzmann equation, treating each particle as a discrete element of the phase-space distribution of the DM halo---see Ref.~\cite{Rocha2013} Appendix A for a full derivation. Briefly, a particle $i$ at position $\mathbf{r}$ is associated with the density kernel $W(r,h_i)$ with ``smoothing length'' $h_i$, which can be thought of as the radius over which the particle is smeared to give it some non-zero volume. The distance $h_i$ must be chosen carefully; setting it too large leads to non-local interactions, while setting it smaller than the mean particle spacing leads to non-physical, two-body relaxation effects. Throughout, we use $h_{s,i}$ to refer to the smoothing length for the self interactions, distinguishing it from $h_{g,i}$, the gravitational smoothing length (discussed in more detail in Sec.~\ref{sec:forcesoft}).

In this framework, the scattering rate of some particle $i$ from a target particle $j$ is given by
\begin{equation}\label{Gizrate}
    \Gamma_{ij} = (\sigma/m) \, m_{\rm p} \, v_{\rm rel} \, g_{ij} \, ,
\end{equation}
where $m_{\rm p}$ is the mass of the DM simulation particle, $v_{\rm rel} = |\mathbf{v}_i - \mathbf{v}_j|$ is the relative velocity of the two particles,  and $g_{ij}$ is a number-density factor derived from the density kernel of the respective particles:  
\begin{equation}\label{gij}
    g_{ij} = \int d^3\textbf{x} \, W\left(|\textbf{x}|,h_{s,i}\right) \, W\left(|\textbf{x}+\delta \textbf{x}_{ij}|,h_{s,j}\right) \, ,
\end{equation}
where $h_{s,i(j)}$ is the self-interaction smoothing length for the $i^{\rm th}(j^{\rm th})$ particle. 
The angular integral is taken over the entire volume of the kernel and $\delta \textbf{x}_{ij}$ is the distance between the two particles. $W(r,h_i)$ is generally chosen to be the cubic-spline kernel:\footnote{To ensure the resulting scattering probability matches Eq.~\ref{Gizrate}, the kernel is normalized such that  $4 \pi \int^h_0 dx \, x^2 g_{ij}(x) = 1$. Consequently, the overlap factor: $g_{ij} = [0,1]$}
\begin{equation}
        W\left(r,h_i\right) = \frac{8}{\pi h_i^3}\begin{cases} 
          1-6\left(\frac{r}{h_i}\right)^2 + 6\left(\frac{r}{h_i}\right)^3 & 0 \leq \frac{r}{h_i} \leq \frac{1}{2}\\
          2 \left(1-\frac{r}{h_i}\right)^3 & \frac{1}{2} < \frac{r}{h_i} \leq 1\\
          0 & \frac{r}{h_i} > 1 \, .
\end{cases}
\end{equation}
In principle, other kernel forms could be used here, which may affect the simulation results. As such, our results pertain specifically to this choice of the kernel. GIZMO approximates the integral in Eq.~\ref{gij} by taking the average  of the particle smoothing lengths and treating the result as a constant length, $h_{\rm avg} = \left(h_{s,i} + h_{s,j}\right)/2$, in the kernel expression. With this simplification, a table of $g_{ij}(\delta x_{ij})$ can be generated at the start of a simulation and the integral is simply a function of $\delta x_{ij}/h_{\rm avg}$. 

Given Eq.~\eqref{Gizrate}, the probability that particle $i$ scatters in a time step $dt_i$ is
\begin{equation}
     P_{ij} = \Gamma_{ij} \, dt_i
\end{equation}
with the total probability of interaction between the particles being
\begin{equation}
    P_{ij} = P_{ji} = \frac{P(i|j) + P(j|i)}{2} \, .
\end{equation}
Whether the pair of particles actually scatter is determined by drawing a random number $R \in [0,1]$ and comparing it to the probability. If $R < P_{ij}$, then a kick is applied to both  particles in the center-of-mass frame. The post-interaction velocities are:
\begin{equation}
\label{eq:kick}
    \textbf{v}_i' = \textbf{v}_c + \frac{m_j}{m_j+m_i} v_{\rm rel} \ \mathbf{\hat{e}}  \quad \quad \quad 
   	\textbf{v}_j' = \textbf{v}_c - \frac{m_i}{m_i+m_j} v_{\rm rel} \ \mathbf{\hat{e}} \, ,
\end{equation}
where $\textbf{v}_c$ is the center-of-mass velocity and $\mathbf{\hat{e}}$ is a random direction.

\subsubsection{Spline Method}

The Spline method, based on the approach described in Ref.~\cite{Vogelsberger2012}, is unique because the scattering is not determined on a pair-by-pair basis. The total probability of scattering $P_{i}$ is calculated first and then a neighbor is chosen to scatter with. This total probability is built up from the individual interaction probabilities between two particles:
\begin{equation}
    P_{ij} = (\sigma/m) \ W(\delta x_{ij},h_{s,i}) \ m_{\rm p} \ v_{\rm rel} \ dt_i \, ,
\end{equation}
where the cubic-spline kernel is taken for $W(r,h_{s,i})$. For a given particle $i$, the total probability of scattering is the sum
\begin{equation}
    P_i = \sum_{j=0}^{N} \frac{P_{ij}}{2}=\sum_{j=0}^{N} (\sigma/m) \ W(\delta x_{ij},h_{s,i}) \ \frac{m_{\rm p}}{2} \ v_{\rm rel} \ dt_i \,,
\end{equation}
where $N$ is the discreet number of neighbors within the kernel length, and the factor of two in the denominator arises because two particles participate in a scattering event.
A collision occurs if $R < P_i$, for some uniform random number $R \in [0,1]$.  To select the nearest neighbor that participates in the scattering event, all the nearby particles are ranked by their distance to $i$. The target for the collision is chosen as the first particle $l$ that satisfies $R < \sum_j^l P_{ij}$. A velocity kick is then applied following Eq.~\ref{eq:kick}.

\subsubsection{Top-Hat Method}

The third DM collision method considered here was first introduced in Ref.~\cite{Kochanek2000}. It differs from the previous two approaches because it uses a top-hat rather than a cubic-spline kernel. As such, there is no explicit weighting of the scattering probability by the particle separation. In this case, the probability of scattering between a pair of particles is 

\begin{equation}
    P_{ij} = \frac{(\sigma/m) \, m_{\rm p} \ v_{\rm rel} \, dt_{i}}{\frac{4}{3}\pi h_{s,i}^3} \,.
\end{equation}
 A DM-DM scattering event occurs if $R < P_{ij}$ at a given time step, where $R\in [0,1]$ is a uniform random number. The final particle kinematics is again set by Eq.~\ref{eq:kick}. 

\subsection{Implementation of Force-Softening Scheme }
\label{sec:forcesoft}

The choice of force-softening length, $h_i$, plays a key role in determining the robustness of the SIDM halo evolution. If the force softening is adaptive, the scale $h_i$ is determined by
\begin{equation}\label{h_adapt}
    \frac{4\pi}{3}h^3_i \sum_{j=1}^N W(\delta x_{ {ij}},h_{i}) = N_{\rm eff} \, , 
\end{equation}
where $N_{\rm eff}$ is the effective number of neighbors~\citep{Springel2005,Hopkins2015,Hopkins2018}. We set a minimum softening of $h_i = 3$~pc for all adaptive runs. 
For the cubic-spline kernel, $N_{\rm eff} = 32$ is the standard choice in GIZMO, providing a good balance between computational expense and accuracy~\citep{Hopkins2018}.

In this work, adaptive softening is always used for self interactions and is the default for gravitational interactions ($h_i = h_{g,i} = h_{s,i}$ in Eq.~\eqref{h_adapt}). However, Sec.~\ref{sec:gravsoft} also explores the effect of using a fixed gravitational softening length for the low-concentration halo. The fixed force softening (Plummer equivalent $h_{g,i} = 2.8 \epsilon$ for cubic spline) is determined using the criteria of Ref.~\cite{Zeng2022}, which is based on the constraints previously described in Ref.~\cite{vdBosch2018}:
\begin{equation}
    \label{eq:epsilon}
    \epsilon = r_{\rm s} \left[ \ln\left(1+c_{\scriptstyle{200}}\right) -\frac{c_{\scriptstyle{200}}}{1+c_{\scriptstyle{200}}}\right] \sqrt{\frac{0.32 \, \left(N_{\rm p}/1000 \right)^{-0.8}}{1.12 \, c_{\scriptstyle{200}}^{1.26}}} \, .
\end{equation}
This results in softening values much smaller than the criterion of Ref.~\cite{Power2003}, but comparable to values determined by the adaptive softening algorithm, i.e. $\epsilon \sim10 \rm~pc$. In general, the gravitational softening length should have minimal direct impact on the halo evolution as it is at least several times smaller than the core size. For the low-concentration halo, Eq.~\ref{eq:epsilon} yields $\epsilon = 12.6~{\rm pc}$, which is more than an order-of-magnitude smaller than the core size until far into core collapse.

\subsection{Implementation of the Time-Stepping Criterion}
\label{sec:timestep_implementation}

The time step, $dt_i$, is a key  parameter to set when initializing an SIDM simulation run. 
 Considerations of a particle's gravitational acceleration motivates setting the time step of particle $i$ as
\begin{equation}
    d t_{i} = \sqrt{\frac{2\eta h_{g,i} }{a}} \,,
    \label{eq:eta}
\end{equation}
where $a$ is the magnitude of the acceleration that the particle experiences. The tolerance parameter, $\eta$, is a dimensionless number that describes the fraction of the force-softening length that the particle is allowed to move in the given time step. The default value in GIZMO is $\eta = 0.02$ \citep{Hopkins2015}. In this work, we use $\eta = 0.02$ as the baseline scenario, but also consider $\eta = 0.002$.

However, one must also take into account the number of scattering events that occur between two DM particles in a single simulation time step. If a particle scatters multiple times in $dt_i$, and on different CPUs, energy and momentum may not be conserved. Therefore, $dt_i$ should be small enough that the probability of multiple scatters is itself small.
 
 For both the Kernel-Overlap and Top-Hat methods, a time-stepping criterion is applied that requires $P_{ij} < \kappa$, where $\kappa$ is the maximum probability of scattering for a pair of particles.\footnote{For comparison purposes, it is useful to implement a formalism for which the time steps across SIDM implementations are equivalent for a given choice of $\kappa$.  For this reason, we use Eq.~\eqref{eq:kappa1} for both Kernel Overlap and Top Hat.} In practice, this means that 
\begin{equation}
    \Gamma_{ij}(\sigma/m ,h_{\rm avg}) \, dt_i < \kappa \, .
    \label{eq:kappa1}
\end{equation}
We consider two possible limits for the time-stepping criterion: $\kappa = 0.02$ and $0.002$. By default, $\kappa = 0.2$ ~\citep{Rocha2013} in GIZMO and $\kappa = 0.02$ in Arepo~\citep{Vogelsberger2012}. As will be discussed in Sec.~\ref{sec:timestep}, the difference between a 2\% and 0.2\% probability is enough to generate noticeable differences in the late-time evolution of an SIDM halo.

The time-stepping criterion will be different for the Spline method, given that the probability that particle $i$ scatters depends on its nearest neighbors.  In this case, the constraint becomes
\begin{equation}\label{11}
  \sigma \, \rho_{\rm loc} \, v_{\rm loc} \, dt_i < \kappa \, ,
\end{equation}
where $v_{\rm loc}$ is the local velocity dispersion, which is estimated as the maximum relative velocity between a neighbor and particle $i$---see Refs.~\cite{Springel2005,Hopkins2015}---and the local density is
\begin{equation}\label{denW}
    \rho_{\rm loc} = \sum_{j=0}^{N} m_{\rm p} \, W\left(\delta x_{ij},h_{s,i}\right)  \, .
\end{equation}
 In practice, the value chosen for $dt_i$ is the minimum of the two estimated with Eq.~\ref{eq:eta} and Eq.~\ref{eq:kappa1} (or Eq.~\ref{11}).  This ensures that the time stepping is small enough to address concerns regarding both the gravitational and self interactions.

\subsection{Characterization of the Halo Core}

For the purposes of this study, it is necessary to characterize halo properties such as the central density and core size. To recover a density profile for a halo, the location of its center must be known. To determine this, we calculate the center-of-mass of the particles with the highest local densities, as determined by the estimation of $\rho_{\rm loc}$ in Eq.~\eqref{denW} and assuming a cubic-spline kernel. The number of particles used for this evaluation is resolution-dependent. For the low-, medium- and high-resolution simulations, we choose 200, 3000, and 6000 particles, respectively.
In general, the center-of-mass of the entire halo does not coincide with the center-of-mass of the halo's core because, over time, the core shifts relative to the outer regions of the halo, especially for the low-resolution simulations. Using the core's center-of-mass allows for a more reliable determination of the core density and velocity dispersion profiles. 

Once the core's center-of-mass is determined, the particles in the halo are divided into 100 evenly-spaced radial logarithmic bins for which the density and velocity dispersion are obtained. To quantify the core density and size, the following density profile is fit to the inner region of the halo,
\begin{equation}
    \rho (r) = \frac{\rho_{\rm core} }{\left(1 + (r / r_0)^2 \right)^{3/2}} \, ,
\end{equation}
where $\rho_{\rm core}$ is the core density and $r_0$ is the characteristic radius beyond which the slope of the log profile transitions from a constant to $-3$. This density profile is a good approximation to the isothermal density profile in the inner regions. The fitting is done with a non-linear least-square method. However, the outer slope of this profile will not hold far outside of the core, where the slope of the log profile transitions to $-2$. For this reason, we determine appropriate fitting boundaries at every simulation snapshot. First, the central density $\rho_{\rm core}$ is estimated by averaging the local densities of the 200 central-most particles and then the profile is fit out to the radius where the density drops to $\rho_{\rm core}/5$.
We define the core radius, $r_{\rm core}$, as the radius where the density drops to half the core density, which implies $r_{\rm core} = r_0 \sqrt{2^{-2/3}-1}$. 
Given the expectation that the core is isothermal throughout the halo's evolution, we compute the instantaneous core velocity dispersion using all the particles within $r_{\rm core}$ and relate the 3D and 1D dispersions assuming an isotropic velocity distribution, $v_{\rm core, 3D}^2 = 3v_{\rm core}^2$.

\section{Numerical Effects on SIDM Halo Evolution}
\label{sec:numerical}

This section explores how various numerical implementations of an N-body code affect the evolution of an SIDM halo. In particular, we consider effects of the SIDM implementation method (Kernel Overlap, Spline, or Top Hat), the numerical resolution, the gravitational force softening, and the time-stepping criterion. Each of the following subsections examines the effects of these variations on outputs such as the density profile evolution or the total energy of the system.

\subsection{SIDM Methods}
\label{sec:highres}

\begin{figure*}[ht]
    \centering
    \includegraphics[width=\textwidth]{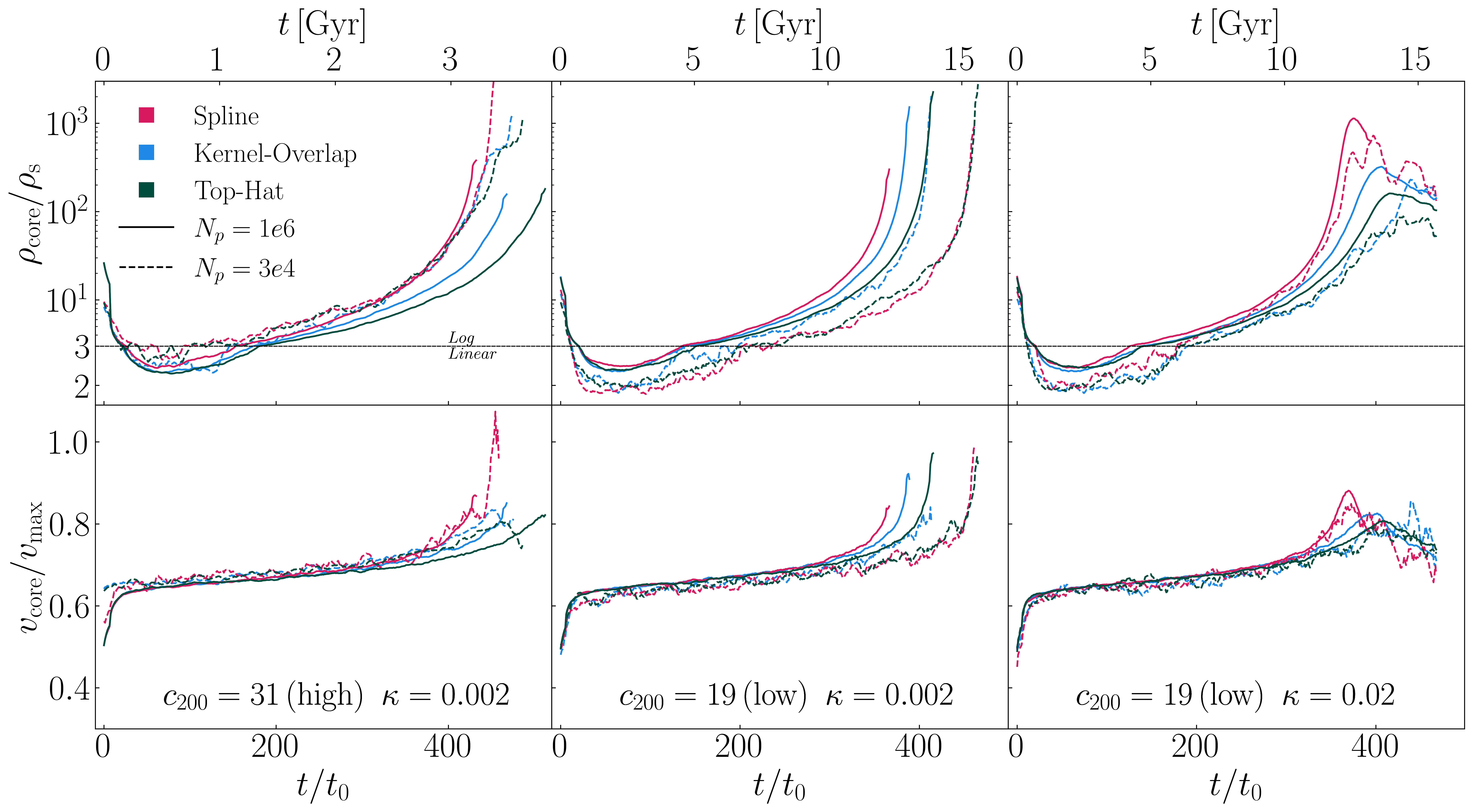}
    %\captionsetup{labelformat=empty}
    \caption{The evolution of a $1.15\times 10^9~M_\odot$ halo with either a $c_{\scriptstyle{200}} = 19$ or $31$ concentration, assuming a self-interaction cross section of $\sigma/m = 50~\text{cm}^2/\text{g}$. The low-concentration halo, which collapses late in cosmological time, is generated for both $\kappa = 0.002$ and $0.02$.  The high-concentration halo, which collapses earlier, is only generated for $\kappa = 0.002$. The halo evolution is shown as a function of dimensionless time, normalized in terms of the thermal relaxation time of the halo, $t_0$. The top axis provides the dimensionful time in Gyr.  \textbf{Top:} The evolution of core density, $\rho_{\rm core}$, normalized to the NFW scale radius, $\rho_{\rm s}$.  Note the two vertical axes: a linear~(log) scale is used for densities below~(above) $3\rho_{\rm s}$.
    Results for the Spline, Kernel-Overlap, and Top-Hat SIDM implementations are shown in red, blue, and green, respectively.  The solid~(dashed) lines correspond to the high~(low)-resolution simulations. 
\textbf{Bottom:} The evolution of velocity dispersion, $v_{\rm core}$, normalized to the halo's maximum velocity, $v_{\rm max}$.  All plotted curves have been smoothed by averaging over the six nearest snapshots, which span about $10t_0$, with the exception of very early or late times where the halo evolution is rapid and no smoothing is necessary. For the high-resolution simulations, all three SIDM methods predict similar evolution during core formation, but diverge during core collapse. Specifically, the Spline method leads to the most rapid halo evolution, while the Top-Hat method leads to the slowest. The low-resolution simulations show considerable variation in the minimum density, which ultimately affects the core-collapse timescale. As discussed in Sec.~\ref{sec:lowres}, this is traced back to numerical noise in the initial conditions. Lastly, simulations with $\kappa = 0.02$ experience a failure at late times where core-collapse reverses, regardless of the SIDM method and resolution. These results pertain specifically to halos simulated with adaptive force softening for gravity.}
    \label{fig:evoall}
\end{figure*}

One of the main goals of this study is to compare between different SIDM implementation methods. Fig.~\ref{fig:evoall} shows the core density normalized to the NFW scale density, $\rho_{\rm core}/\rho_{\rm s}$ (top panels), and core velocity dispersion normalized to the maximal circular velocity of the halo, $v_{\rm core}/v_{\rm max}$ (bottom panels), for the Spline~(red), Kernel-Overlap~(blue) and Top-Hat~(green) methods. For each curve, time is plotted in units of the thermal relaxation timescale of the halo, $t/t_0$, where $t_0 \equiv (\sqrt{16/\pi} \rho_{\rm s} v_0 \sigma/m)^{-1}$ and $v_0 \equiv \sqrt{4\pi G \rho_{\rm s} r_{\rm s}^2}$~\citep{Balberg2002, Nishikawa2020}. Shown on the same plot are variations on the number of particles, the time-stepping criterion, and the concentration of the halo used for the initial conditions of the simulation. The effects of some of these variations will be discussed below.

For all cases plotted in Fig.~\ref{fig:evoall}, the general behavior of the density evolution follows the characteristic trajectory expected from the SIDM fluid model~\citep{Balberg2002,Koda2011,Essig2019,Nishikawa2020,Outmezguine2022,Yang2023,Jiang2023}. In the initial stages of halo evolution, heat flows inwards and the core grows in size until it reaches a maximum radius, corresponding to some minimum central density. At this point, the central region of the halo is much hotter than its surroundings and eventually heat flow reverses direction, triggering the onset of runaway core collapse. During this final stage, the core rapidly shrinks in size and increases in both temperature and density. However, as is evident from Fig.~\ref{fig:evoall}, the detailed features of this evolution can vary significantly depending on the numerical implementation.

A general finding is that during the core-expansion phase, the core temperature is more robust than the core density to the different implementations and resolutions and always gives results that are close to the expectation from the fluid approach to within about $2\% v_{\rm max}$ (this will be further discussed below). The different implementations do significantly affect the core density during core expansion, the general trend being that simulations for which the minimum core density is larger collapse faster (and vise-versa). Thus, capturing physical SIDM behavior at early times in a halo's evolution significantly affects the late-time behavior.

In all high-resolution simulations, the Spline method produces the earliest core collapse, followed by the Kernel-Overlap and then the Top-Hat method. To quantify this comparison, one can define the collapse time, $t_{\rm coll}$, as the time at which the core density reaches 100 times the scale density, $\rho_{\rm core}(t_{\rm coll}) \equiv 100 \rho_{\rm s}$ (evolving halos to higher core densities involves significantly higher computational cost).   For the high-concentration halos, the collapse time is $t_{\rm coll} \approx 416 t_0, 461 t_0,$ and $500 t_0$ for the Spline, Kernel-Overlap, and Top-Hat method, respectively.  For the low-concentration halos simulated with $\kappa = 0.002$, $t_{\rm coll} \approx (338$--$394)t_0$, depending on the SIDM method. Holding all other variables constant, the higher $\kappa = 0.02$ simulations tend to collapse $\sim2\%$ faster. Additionally, for some cases, the behavior of both central density and velocity dispersion stall and become non-physical at late times.

One possible cause for the differences in halo evolution is the scattering rate of particles within the simulation. Theoretically, the rate is expected to be,
\begin{equation}
    \Gamma_{\rm exp} = \int_V \frac{\rho(\mathbf{x})^2}{2m_{\rm p}^2} \, \langle \sigma v_{\rm rel} \rangle \, dV \,,
\end{equation}
where $\rho(\mathbf{x})$ is the halo's density profile and $\langle \sigma v_{\rm rel} \rangle$ is the thermal average of the cross section times the relative velocity. The thermal averaging can be calculated by assuming that all interacting particles have a Maxwell-Boltzmann distribution of the form $f(\mathbf{v}) = [1/(2 \pi v_r^2)]^{3/2} e^{-(\mathbf{v}/v_r)^2/2}$ (where $v_r(\mathbf{x})$ is the position-dependent 1D radial velocity dispersion of the halo), which for a constant cross section just gives $\langle \sigma v_{\rm rel} \rangle = 4/\sqrt{\pi} \times \sigma v_r$. To estimate $\Gamma_{\rm exp}$, we spline the density and dispersion profiles for each simulation snapshot, which is taken every $t_0$ to capture the halo's evolution during core collapse, and numerically integrate over the entire simulation volume. The expected scattering rate can then be compared to the scattering rate observed in the simulations. To obtain this rate, we count the number of scattering events, $N_{\rm scat}$, within each time interval, $t_0$, in the entire simulation volume, and take $\Gamma_{\rm obs} = N_{\rm scat}/t_0$.
\begin{figure}
    \centering
    \includegraphics[width=3.5in]{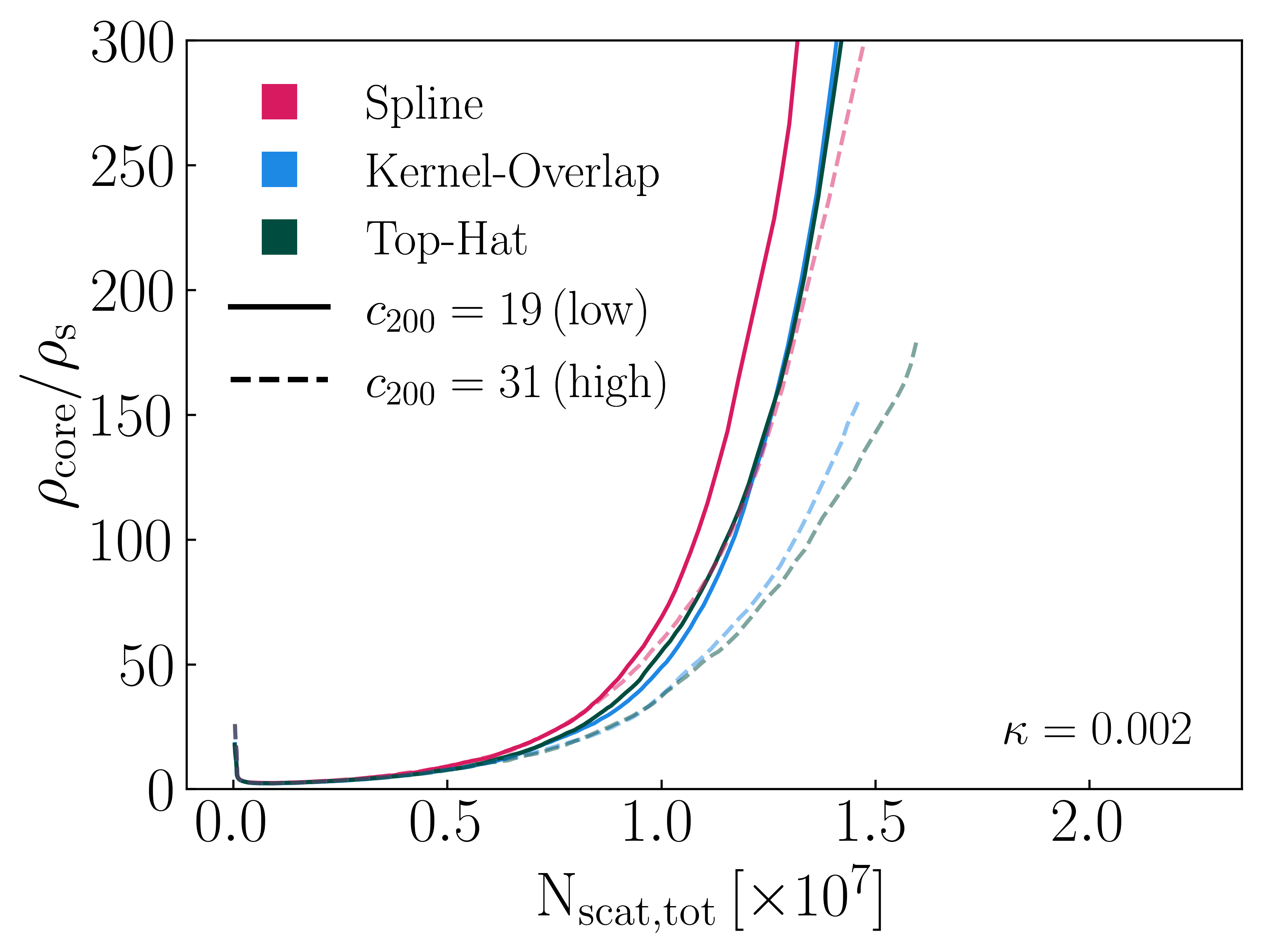}
    \caption{The evolution of core density as a function of the total scattered mass for the high-resolution simulations with $\kappa= 0.002$. The solid opaque lines correspond to the low-concentration halo, while the dashed  lines are for the high-concentration halo. The Kernel-Overlap and Top-Hat methods achieve the same core density at equivalent scattered mass, while the Spline method achieves the same core density with less scattering. These results pertain specifically to halos simulated with adaptive force softening for gravity.}
    \label{fig:rhoc/t.r}
\end{figure}

For all simulations run in this study, $\Gamma_{\rm exp}$ and $\Gamma_{\rm obs}$ agree with each other to within about $10\%$ and are remarkably constant in time. In particular, for the highest-resolution simulations, the ratios are constant throughout the entire evolution with the exception of the last few snapshots that are deep in the core-collapse regime (when $\Gamma_{\rm obs}$ is extremely sensitive to the rapid evolution). These ratios are $\Gamma_{\rm obs}/\Gamma_{\rm exp} \approx 1.08, 1.06,$ and $1.02$ for the Kernel-Overlap, Spline, and Top-Hat implementations, respectively (see Fig.~\ref{fig:r/r_e} for additional details). In this regard, we note that Ref.~\cite{Meskhidze2022} found larger differences in the scattering rates between the Arepo and GIZMO codes, and these differences were cross-section dependent. We are unable to make a direct comparison to these results because GIZMO and Arepo are different in more ways than just the SIDM implementations. Additionally, the SIDM kernels we use are adaptive, while Ref.~\cite{Meskhidze2022} used a fixed softening for SIDM that was a fraction of the gravitational force softening (which was also not adaptive).

To study whether the different scattering rates contribute to the variations in evolution for different SIDM implementations, one could think of the cumulative number of scattering events at any given time in the simulation as a universal clock. Namely, since different SIDM implementations have different instantaneous scattering rates, it is plausible that evolution would be equivalent if plotted as a function of the cumulative number of scattering events instead of actual simulation time. To test this, the total number of scattering events is computed, $N_{\rm scat,tot}(t) \equiv \sum_{t_i=0}^t N_{\rm scat}(t_i)$ (where $t_i$ denotes the time intervals between the first snapshot and any snapshot at time $t$). These values differ slightly between the different SIDM implementations. For example, at the time of maximal core (approximately $\sim 70 t_0$), $N_{\rm scat,tot}(70 t_0) = 9.40 \times 10^5, \, 9.25 \times 10^5$ and $8.69 \times 10^5$ for the Kernel-Overlap, Spline, and Top-Hat implementations, respectively. Halo evolution as a function of $N_{\rm scat,tot}$ is plotted in Fig.~\ref{fig:rhoc/t.r} for the highest-resolution runs and for both high- and low-concentration simulations with $\kappa = 0.002$. Using this variable instead of time brings the Kernel-Overlap and Top-Hat methods into close agreement, while the Spline method still collapses earlier than the others. This suggests that, while differences in the scattering rates between the SIDM implementations could explain some of the variations in collapse times, additional factors must also be affecting the results. In what follows, we study some additional aspects of core-collapsing SIDM simulations.

\subsection{Mass Resolution and Initial Conditions}
\label{sec:lowres}
\begin{figure}
    \centering
    \includegraphics[width=\textwidth]{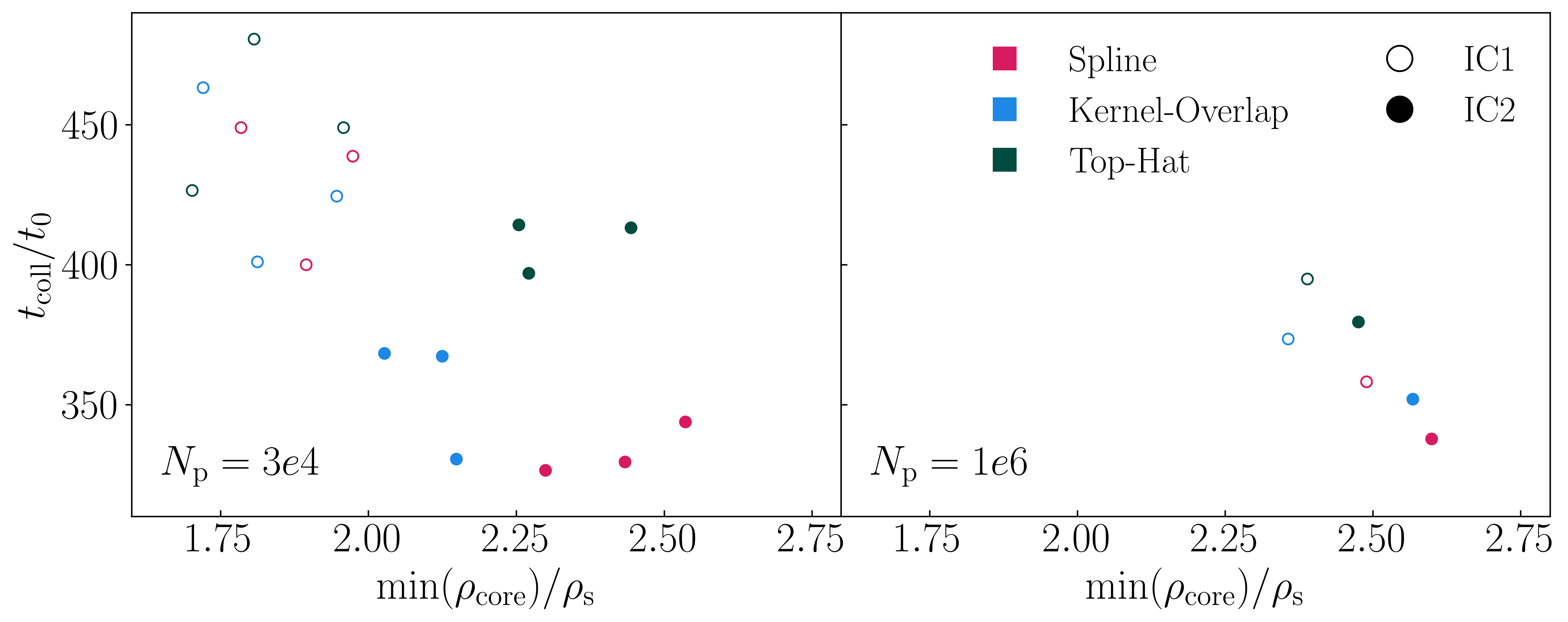}
    \caption{The collapse time, $t_{\rm coll}$, in units of relaxation time plotted as a function of minimum core density, normalized to $\rho_{\rm s}$. Results are shown for the low-~(left panel) and high-resolution~(right panel) simulation of the low-concentration halo with $\kappa = 0.002$. They are plotted for two different realizations of the NFW initial conditions of the halo (IC1 and IC2), shown by the empty/filled circles. For a given initial condition and SIDM method (indicated by color), each low-resolution halo is re-simulated three times to test reproducibility. The dominant source of scatter in the minimum core density and collapse time for the low-resolution simulations is the numerical noise in the generation of the initial conditions. These results pertain specifically to halos simulated with adaptive force softening for gravity. }
    \label{fig:tcollpcore}
\end{figure}

The total number of particles in a halo must be large enough to ensure that halo properties are well resolved and that numerical noise does not have an effect on the halo's evolution. This subsection explores the effects of reducing the particle count below $10^{6}$ per halo. The most dramatic differences occur for the lowest-resolution halos in the suite, the results of which are indicated by the dashed lines in Fig.~\ref{fig:evoall} (curves for medium-resolution simulations are not shown because they are nearly identical to those of the high-resolution simulations).  These low-resolution halos exhibit significant variation in both the minimum core density as well as the core-collapse time.   
For example, the low-concentration halo has $\text{min}\left(\rho_{\rm core}\right) \approx (1.6$--$2.0) \rho_{s}$ for the low-resolution runs, while the range narrows to $\approx (2.3$--$2.5) \rho_{s}$ for the high-resolution runs (see Fig.~\ref{fig:tcollpcore}). 

In general, for cases where the cores are larger and less dense, the low-resolution halos take longer to core collapse than their higher-resolution counterparts. For the low-concentration halo simulated with $\kappa=0.002$, the core-collapse times are $t_{\rm coll} \approx 450 t_0 $ for the Spline and Top-Hat method and $t_{\rm coll} \approx 400 t_0 $ for the Kernel-Overlap case. As mentioned above, the early-evolution of the halos ultimately impacts its late-time evolution. These effects are due to numerical noise in the initial conditions and are not physical.

To further test numerical noise, Fig.~\ref{fig:tcollpcore} plots the collapse time and minimum core density for multiple realizations of the low-concentration halo with $\kappa = 0.002$. The left panel corresponds to low resolution and the right panel to high resolution. In each case, open circles correspond to the baseline NFW initial conditions used in this work~(IC1) while the filled circles correspond to a separate, independent set of initial conditions~(IC2). For the low-resolution case only, and for a given initial condition and SIDM implementation, each halo is simulated three times to test whether additional numerical effects, such as issues with random seed generators, are at play.  This is not done for the high-resolution simulations due to computational costs.

As the left panel of Fig.~\ref{fig:tcollpcore} demonstrates, there are at least two sources of variability for the low-resolution simulations. A dominant source of uncertainty is related to the choice of initial conditions, which leads to variability of order $30\%$ in both collapse time and minimal core density. A sub-dominant source of uncertainty, which is likely due to random number seed generation in the code (although we cannot isolate the effect explicitly), leads to $\sim~10\%$ variations in collapse time and $\sim 5\%$ variations in minimal central density. The right panel shows results for the higher-resolution simulations. Clearly, the spread in these results, corresponding to varying initial conditions, is now much smaller, of order $10\%$ for both collapse time and minimal core density.

As discussed earlier, the minimal core density and collapse times are in fact related to each other since larger minimal core densities generally correspond to earlier collapse times. This is somewhat expected since the instantaneous collision timescale in the core is proportional to $(\rho_{\rm core} v_{\rm core})^{-1}$. Numerically, we find that the values of $v_{\rm core}$ at the time of maximal core are approximately constant when varying SIDM methods and initial conditions, and therefore one might expect that $t_{\rm coll} \propto \text{min}(\rho_{\rm core})^{-1}$ which is approximately the scaling observed in Fig.~\ref{fig:tcollpcore}.

Given the results of this section, from this point onward, the analyses focus solely on the high-resolution simulations.

\subsection{Time-Stepping Criterion} 
\label{sec:timestep}

\begin{figure*}
    \centering
    \includegraphics[width=\textwidth]{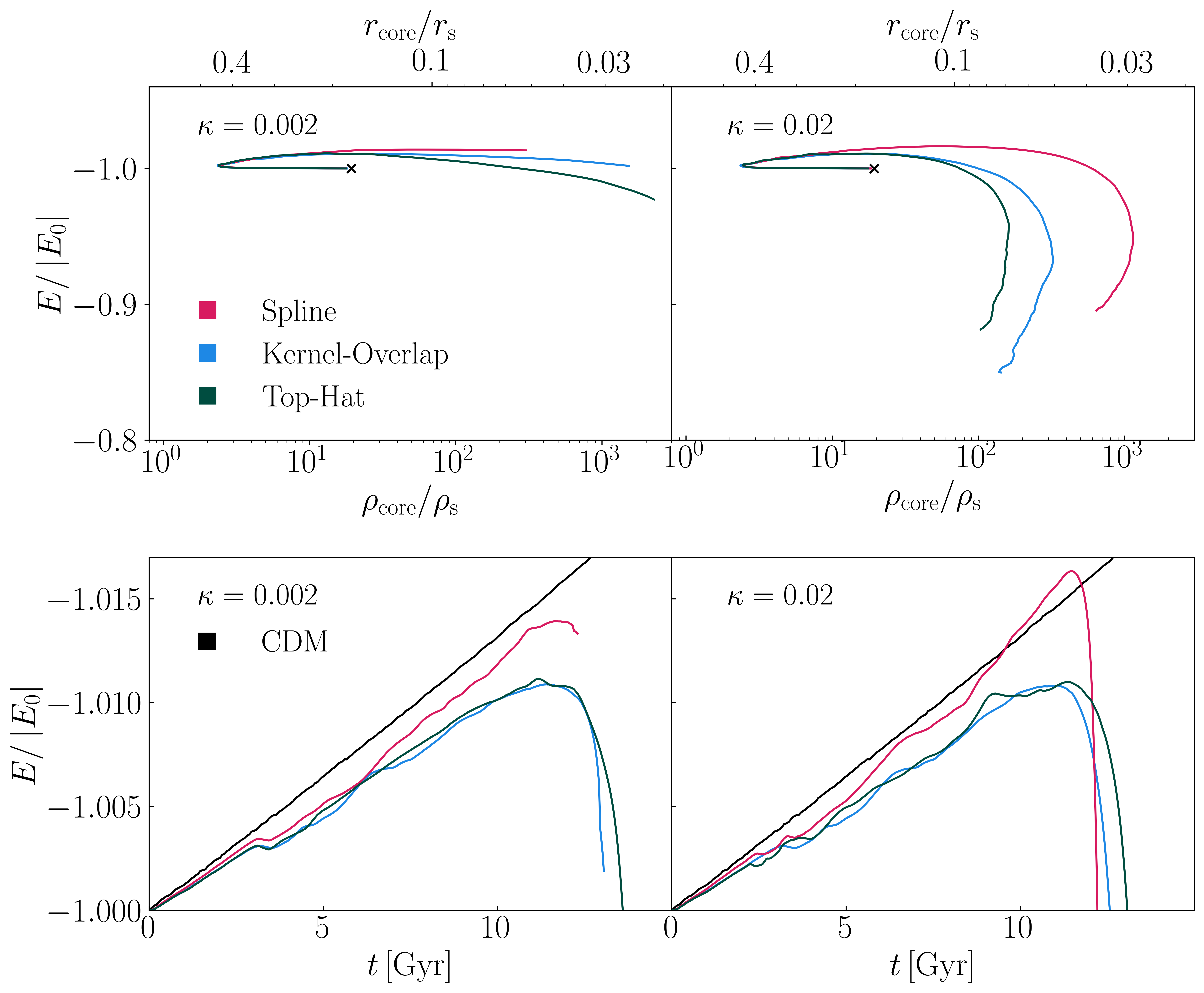}
    \caption{Top: The evolution of total energy versus core density for the low-concentration halo with $\kappa=0.002$ and $0.02$. Plotted in this space, the halo's energy evolves clockwise with time. This evolution is characterized by a phase of core expansion at constant energy followed by the halo's energy becoming more negative, i.e. losing thermal energy, from the time of core formation to core collapse. Simulations with $\kappa = 0.02$ exhibit a critical failure of energy conservation once core densities reach 50--500$\rho_{\rm s}$ (depending on SIDM method), where both the kinetic and potential energy is rapidly lost. Generally, the Spline method experiences this failure at much higher core densities than the other two methods.
    Bottom:~The evolution of halo energy over time. For comparison, the evolution of a CDM halo is plotted in black. The Spline method produces energy evolution closest to the CDM halo at times past the ``maximum core'', showing that this method produces the least net energy gain due to the DM scattering. These results pertain specifically to halos simulated with adaptive force softening.}
    \label{fig:energy}
\end{figure*}

Energy non-conservation due to numerical processes plays a key role in SIDM halo evolution. The top panel of Fig.~\ref{fig:energy} plots the evolution of the total energy,\footnote{For the following discussion, ``total energy'' refers to the total potential and kinetic energy of all the particles in the simulation.} normalized to the initial energy, $E_0$, of each simulated halo as a function of core density. As the halo evolves in time, it moves clockwise along each curve, starting from the initial state (marked with an `x'), then moving towards lower core density through core expansion, and then towards higher core density in the subsequent core collapse. 
For the $\kappa=0.002$ example (top left), the energy is well-conserved ($E/|E_0| \approx -1$) while $\rho_{\rm core} \lesssim 10 \rho_{\rm s}$. Far into the core-collapse regime, though, the different SIDM implementations result in different energy evolution. Simulations run with the Spline method show the largest energy loss, while the Top-Hat simulations exhibit the least energy loss. At very late times, some of the halos start to exhibit significant gains in energy.

This behavior is even more apparent in the bottom row of Fig.~\ref{fig:energy}, which shows the energy evolution as a function of time. Changes in energy are driven by two distinct numerical effects that compete with each other.  The first is energy loss from the implementation of gravitational scattering.  This effect should be present even for a CDM simulation, which is indeed the case, as demonstrated by the solid black lines in the bottom row of Fig.~\ref{fig:energy}.  The second main effect arises from the implementation of the self interactions.  The additional energy gains that result from this effect are apparent in the bottom panel of Fig.~\ref{fig:energy}, where all three SIDM lines differ from the black CDM expectation.  
In the initial stages of core collapse, we observe a $\sim1.0$--$1.3\%$ change in energy for the $\kappa = 0.002$ halo. SIDM offsets the energy loss from gravitational scattering, bringing the halo closer to its initial energy inventory. However, this does not improve the accuracy of the simulations as artificially adding and removing energy throughout the halo influences its thermal evolution unpredictably. 

While these sub-percent changes to the total energy may seem negligible, any numerical energy loss during core collapse is effectively an additional heat source in the central region of the halo where the vast majority of the self interactions take place. As the energy decreases, the potential well of the halo deepens and mass is dragged towards the center of the halo, effectively accelerating core collapse. From Fig.~\ref{fig:energy}, it is clear that the Spline method heats halos the most in the initial stages of core collapse, followed by the Kernel-Overlap and Top-Hat methods. This explains the behavior noted in Fig.~\ref{fig:evoall}, where the Spline method results in the fastest core-collapse times. 

Conversely, any numerical effects that lead to artificial gains in energy will effectively cool the system, delaying core collapse. At very late times in the halo evolution, all the simulations begin to gain a lot of energy, which causes the core to expand again and the entire halo to become less gravitationally bound.  For the $\kappa=0.002$ halo, this cooling does not have a noticeable effect on the core's evolution.  However, for the $\kappa = 0.02$ example, the degree of  numerical cooling is so significant that it causes the core-collapse process to completely reverse. The right-most column of Fig.~\ref{fig:evoall} shows the case of $\kappa=0.02$ for the low-concentration halo. For all three SIDM implementations, the density growth slows down and experiences some fluctuations when the core density exceeds $\sim 200\rho_{\rm s}$. Looking at the evolution of the central velocity dispersion, it is clear that something is going wrong because the dispersion starts to decrease just as the growth of core density begins to fluctuate. Beyond this point, the core collapse halts and reverses with the density dropping by as much as a factor of 10 by the end of the simulation run. At the same time, the central velocity dispersion falls back down to $\sim 0.7v_{\rm max}$, virtually reversing the total heat gain since the start of core collapse. In general, the Top-Hat method fails at lowest core density, while the Spline method fails at the highest. From the right panel of Fig.~\ref{fig:energy}, it is clear that the reversal of core collapse coincides with a rapid turnaround in the halo's $E/|E_0|$. In all three SIDM implementations, there is a decrease of $\sim15\%$ of the total halo energy by the end of the simulation run.

While the energy loss driven by gravitational interactions is tied to adaptive force softening (as discussed in the next section), the energy gains driven by self interactions likely result from particles scattering multiple times in a single time step. Specifically, ''bad events" constitute particle multi-scatters with neighbors that reside on other CPUs, which leads to energy gain since the scattering velocities are not updated between multi-scattering events (see Ref.~\cite{Robertson2017a} for a detailed discussion). Generally the number of bad events is of the same order as the number of multi-scatters because most of the particle neighbors reside on different CPUs. A simple back-of-the-envelope estimate of the double-scattering rate shows these highly infrequent events can generate a significant change in energy even for well-resolved halos. The number of double-scattering events, $N^{(2)}$, in the halo's core is
%\begin{equation}
    %N^{(2)} = \Gamma^{(2)} \Delta t = \int_V P_i^2 n dV dt_i^{-1} \approx \kappa^2 N_{\rm core} \frac{\Delta t} {dt_i},
%\end{equation}
\begin{equation}
    N^{(2)} = N^{(1)} P_i \approx N^{(1)} \kappa \, ,
\end{equation}
where $N^{(1)}$ is the number of single-scattering events and $P_i$ can be approximated as $\kappa$ far into core collapse when the interactions are very frequent. Assuming each double-interaction event adds kinetic energy comparable to a core particle of mass $m_{\rm p}$ with velocity $v_{\rm core}$, the relative change in a halo's kinetic energy is
\begin{equation}
    \frac{\Delta E}{E_{\rm 0, kin}} \approx \frac{N^{(2)} m_{\rm p} v_{\rm core}^2}{M_{\rm halo} v_{\rm core}^2} \approx  \frac{N^{(2)}}{N} \approx \kappa \frac{N^{(1)}}{N} \, ,
\end{equation}
where $E_{\rm 0, kin}$ is the halo's initial kinetic energy, and we approximate the average speed  of a halo particle as being $\mathcal{O}(1) v_{\rm core}$. Clearly, the gain in kinetic energy becomes appreciable as $N^{(2)}$ approaches the total number of particles in the halo, $N$. %Since the ratio of double-to-single scattering events is $N^{(2)}/N^{(1)} \approx \kappa$, we can rewrite this expression in terms of the latter, which we keep track of in the simulation
%\begin{equation}
%    \frac{\Delta E}{E_{\rm i}} = \kappa \frac{N^{(1)}}{N}.
%\end{equation}
For the high-resolution simulations with $N = 10^6$ particles, approximately $N^{(2)} = 10^6$ double-scattering events are needed to change the halo's kinetic energy by $E_{\rm 0, kin}$. During the last $\approx 1~\rm Gyr$ of the simulation concurrent with the rapid changes in the halo's energy, the high-resolution simulations with $\kappa = 0.02$ go through $N^{(1)} \approx \mathcal{O}\left(10^7\right)$ single-scattering events and $N^{(2)} \approx \mathcal{O} \left(10^5\right)$ double-scattering events. This results in an energy change of approximately 
\begin{equation}
    \frac{\Delta E}{E_{\rm 0, kin}} \approx 0.02 \times \frac{10^7}{10^6} \approx 0.1 \, ,
\end{equation}
which is comparable to the energy change during this time of the simulation~($\approx 15 \% E_{\rm 0}$). Since $\Delta E \propto \kappa$, reducing $\kappa$ by an order-of-magnitude greatly improves energy conservation. However, $\kappa = 0.002$ does not, in general, guarantee the absence of artificial heating. Instead, reducing $\kappa$ allows the simulation to reach higher densities, where the energy error per double scattering event increases due to the rising temperature of the core.

%Taking $\kappa = 0.02$ and $N \approx \mathcal{O}(10^4)$, $dt_i \approx \mathcal{O}(10^{-6}) \rm \, Gyr$ for the high resolution simulations results in the total number of double scattering events $N^{(2)} \approx \mathcal{O}(10^6)$ in a span of $1 \rm \, Gyr$ of simulation time. Assuming each double interaction event adds kinetic energy comparable to a core particle with velocity $\approx v_{\rm core} = 0.64 v_{\rm max}$ the resulting change in halo energy is of the same order as halo's initial total energy $\Delta E = 10^6 \times 1/2 m_{\rm p} v_{\rm core}^2 \approx E_{\rm initial}$. Since $\Gamma_i \propto \kappa^2$, reducing $\kappa$ by an order of magnitude should greatly alleviate the issues with energy conservation.}

Refs.~\cite{Robertson2017a,Fischer2021}, implemented several features in their N-body codes to reduce the impact of bad multiple scattering events. In order to reduce the probability of bad scattering, Ref.~\cite{Robertson2017a} enforced a particle communication direction between every pair of the CPUs, while Ref.~\cite{Fischer2021} additionally enforced a communication queue such that a single particle is only actively scattering on a single CPU at a time, which completely eliminates the possibility of simultaneous multi scattering. Not unlike simply lowering the scattering rates in individual particle time steps, this latter approach does come with a significant impact on computational efficiency of the simulation.

A key take-away is that without implementing a numerical scheme that reduces the rate of multi scattering, the time-stepping parameter, $\kappa$, must be chosen judiciously to minimize energy non-conservation. For a given $\kappa$, we find that the Spline method minimizes energy gain or loss compared to the other two methods for the halos simulated here. It is not clear why the Spline method exhibits smaller energy gains, but it could be due to the neighbor sorting when choosing which particle to scatter with, which is a major difference from the other two methods. 

\subsection{Adaptive Gravitational Force Softening} 
\label{sec:gravsoft}

\begin{figure}
    \centering
    \includegraphics[width=\textwidth]{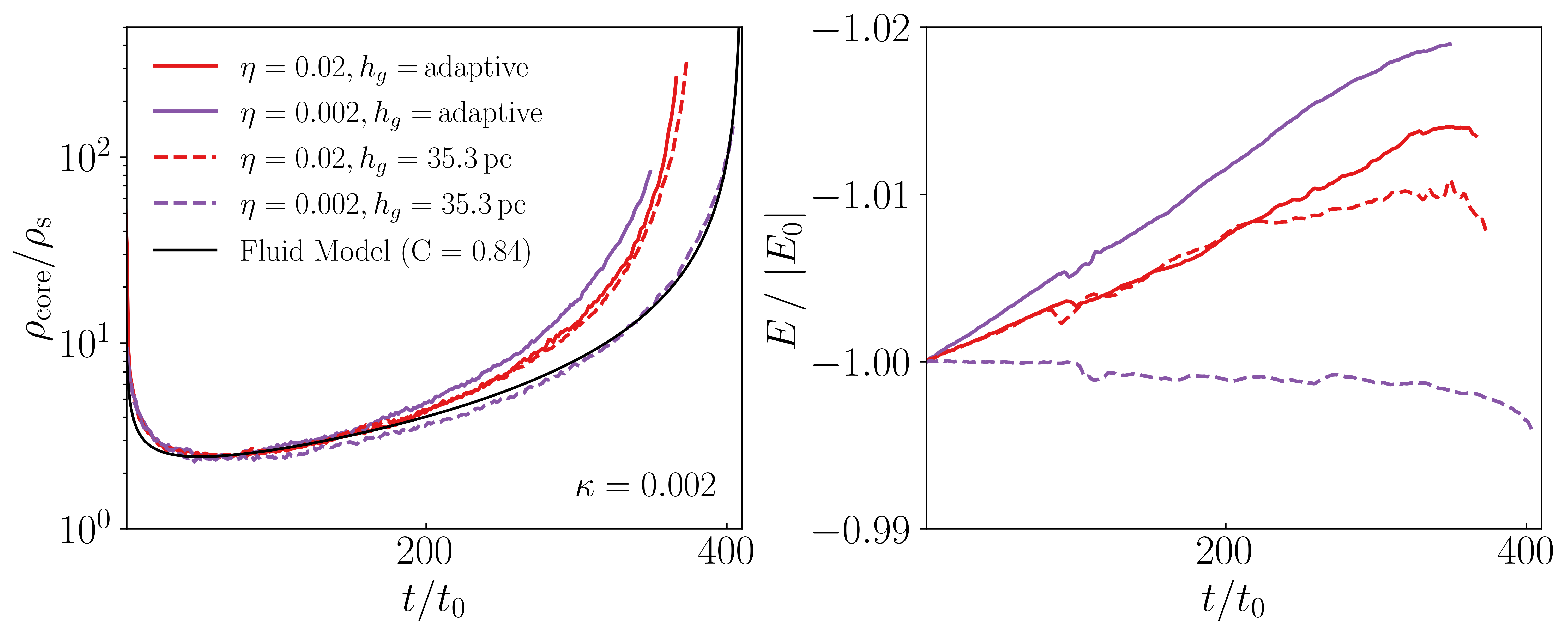}
    \caption{The evolution of core density and energy of the low-concentration halo with fixed and adaptive force softening for gravity, using the Spline implementation. The particle smoothing $h_{s}$ is adaptive. Additionally, the tolerance parameter, $\eta$, is varied. The simulation with fixed force softening and $\eta=0.002$ exhibits much improved energy conservation relative to the others. The black line corresponds to the best-fit fluid model  solution with $C = 0.84$, obtained by fitting to the simulation with best energy conservation.}
    \label{fig:energycons}
\end{figure}

All analyses until this point have focused on simulations that were run using adaptive force softening for the gravitational interactions.  Here, we explore the effects on the halo evolution by instead using a fixed force-softening length, $\epsilon$, and different tolerance parameter, $\eta$. 
Because the goal is to maximize energy conservation, we focus on the Spline method in the following discussion. 
Figure~\ref{fig:energycons} shows the results of re-simulating the low-concentration $\kappa=0.002$ halo for the Spline method. The previously described adaptively-softened simulation is plotted in solid red for comparison. As shown in the right panel of Fig.~\ref{fig:energycons}, reducing $\eta$ while using adaptive softening increases energy loss over the course of the simulation. However, the combination of fixing the force-softening length \emph{and} decreasing the tolerance parameter to $\eta = 0.002$ essentially removes the numerical heating observed in all other simulations. By comparing the evolution of this simulation with the rest, it is clear that the gain of  $E/|E_0| \sim1$--$2\%$ results in roughly a $10\%$ acceleration in core collapse. Because it is unpredictable how much of an effect numerical heating/cooling will have on core evolution in general, adaptive softening for gravity is undesirable for DM particles in core-collapsing simulations.

\section{Comparison to the Fluid Model}
\label{fluid}

The hydrodynamical evolution of SIDM halos is captured by the gravothermal fluid model, extensively explored in several key studies \citep{Lyndel1980,BalbShap2002,Koda2011,Pollack2015,Essig2019,Nishikawa2020,Outmezguine2022,Yang:2022hkm,Yang2023,Jiang2023,Zhong:2023yzk}. The set of partial differential equations describing the evolution of the local density $\rho(r,t)$ at location $r$ and time $t$, velocity dispersion $v(r,t)$ and luminosity $L(r,t)$, are the fluid-momentum equation in the hydrostatic limit, the first law of thermodynamics, and a heat transport equation with an effective heat transport coefficient that is typically used in both the short and long mean-free-path regimes. The heat transport equation is given by
\begin{equation}
    \frac{L}{4 \pi r^2} = -\frac{3}{2} a v \frac{\sigma}{m} \left( \frac{4 \pi G}{C \rho v^2} + \frac{a}{b} \left(\frac{\sigma}{m}\right)^2 \right)^{-1} \frac{\partial v^2}{\partial r} \, ,
    \label{eq:heat_transport}
\end{equation}
where $a = 4/\pi$ and $b = 25 \sqrt{\pi}/32$. For almost all cases of interest, the entire halo initially evolves in the long mean-free-path regime during which the first term in parenthesis dominates. When the core density and velocity dispersion become large, the second term can begin to dominate and the core enters the short mean-free-path regime. Eq.~\eqref{eq:heat_transport} is an approximation because there is no {\it ab initio} derivation of heat transport in the long mean-free-path regime. In that regime, the parameter $C$ has been used as a constant (in space and time) $\mathcal{O}(1)$ fudge-factor in an attempt to calibrate the conductivity using N-body simulations of isolated and idealized halos. Several studies have evaluated this constant and there exist discrepancies of order $25\%$ between different studies with some finding values of $C \approx 0.6$~\citep{Outmezguine2022,Essig2019} while others find larger values of $C \approx 0.75$~\citep{Koda2011} and $C \approx 0.82$~\citep{Yang2023}. 
The various derivations have used different simulation suites with most focused on either the Pippin halo run of Ref.~\cite{Elbert2015} with various cross sections, or isolated halos run by Refs.~\cite{Koda2011} and \cite{Yang2023}. Additionally, the various studies use different methods to fit the fluid model to these simulations. Given the findings of our study, it is plausible that at least some of these discrepancies arise from numerical variations between the simulations.

For completeness, we evaluate our own value of $C$ by comparing the fluid solution of Ref.~\cite{Outmezguine2022}'s run \#1 to our high-concentration, high-resolution, $\kappa, \eta = 0.002$ simulation, with the Spline method and fixed gravitational softening (which exhibits good energy conservation). We choose the value of $C$ for which the fluid method and simulation reach $\rho_{\rm core} = 100 \rho_{\rm s}$ at the same time and find $C \approx 0.84$. The fluid result is shown by the solid black curve in the left panel of Fig.~\ref{fig:energycons}.

After setting $C$, one can then compare the entire simulated halo evolution to the variables that are solved for in the fluid approach. For example, the fluid model predicts that the core density reaches its minimal value at approximately $60 t_0$ for an initial NFW halo in isolation. At that time, the core density is predicted to be $\rho_{\rm core} \approx 2.4 \rho_{\rm s}$, the core radius is $r_{\rm core} \approx 0.45 r_{\rm s}$, and the core velocity dispersion is $v_{\rm core} \approx 0.64 v_{\rm max}$. As shown in Fig.~\ref{fig:evoall}, the results for both the low- and high-concentration simulations are consistent with the fluid predictions. In particular, the minimum core density in the high-resolution simulations\footnote{Given a density profile at every snapshot, we fit a cubic polynomial to $\rho_{\rm core} (t/t_0)$ in the time span $(50$--$200) t_0$ when core collapse is expected to be occurring. From this, we determine the density at maximum core, $\rm min\left(\rho_{\rm core}\right)/\rho_{\rm s}$, by finding the minimum of the resulting curve. This procedure reduces numerical noise in determining these quantities, which can be significant depending on the resolution.}
is $\text{min}\left(\rho_{\rm core}\right) \approx (2.4$--$2.5) \rho_{\rm s}$ and is reached around $t \approx 70 t_0$, regardless of SIDM implementation. There is particularly good agreement in the central velocity dispersion with $v_{\rm core} \approx (0.64$--$0.65) v_{\rm max}$. There are no systematic differences in the core sizes between SIDM implementations,  with $r_{\rm core} \approx 0.42 r_{\rm s}$ for all resolutions simulated, as shown in Fig.~\ref{fig:coresize}. 

\section{Conclusions}
\label{sec:conclusions}

This paper explores the   uncertainties in modeling DM self interactions in N-body simulations, focusing on the regime where the core size is shrinking and the core density is increasing with time. As a concrete case study, we focused on a high- and low-concentration variant of a $1.15\times 10^9~M_\odot$ halo.  We assumed a constant cross section of 50~cm$^2$/g and implemented the self interactions using three common modeling methods found in the literature: the Spline, Kernel-Overlap, and Top-Hat techniques.  Additionally, we also varied the number of particles, time-stepping criteria, and  gravitational force softening used to simulate the halos.
For the isolated dwarf halos considered in this work, we showed that: 
\begin{itemize}
\item Clear differences arise in the halo evolution depending on the SIDM implementation.  Halos evolved using the Spline method collapse fastest, followed by those evolved using the Kernel-Overlap and then the Top-Hat method.  These differences result from the interplay of at least two effects.  First is that the different SIDM implementations yield different scattering rates, which can be between $\sim2$--8\% larger than the expected theoretical value. Second is that the specific numerical implementation of gravitational and DM scattering can result in the halo's total energy not being conserved. Small differences in the energy conservation can translate to large differences in the maximum core densities reached before evolution stalls. We find that the Spline method has the best energy conservation for the halos simulated.
\item An inadequate choice of time step can result in core evolution that stalls or reverses entirely, reducing the observed density and dispersion profiles. Therefore, an appropriate choice of the time-stepping criterion is critical for minimizing the effects of energy non-conservation---e.g., one must ensure that the probability of DM self-scattering in a single time step is very small. For our simulated halos, we found that limiting this probability to 0.2\% ($\kappa = 0.002$) was sufficient to prevent spurious cooling during a halo's core collapse. This choice for $\kappa$ is about an order-of-magnitude lower than the default values used in GIZMO~($\kappa = 0.2$) and Arepo~($\kappa = 0.02$).  
\item Numerical heating that arises from adaptive gravitational softening can accelerate the core-collapse process, even if it only results in a few-percent change in the halo's total energy. We showed  that using a fixed gravitational softening, along with an appropriate choice of tolerance parameter ($\eta = 0.002$), maximized the accuracy of energy conservation for the case of the Spline method.  
\item The evolution of low-resolution halos with $3\times 10^4$ particles is dominated by noise in the generation of initial conditions. In particular, stochastic noise in different realizations of identical $\sim 10^9 M_{\rm \odot}$ NFW halos introduces a $\sim 30$\% scatter in the core density and collapse time.  This is a larger systematic uncertainty than the choice of SIDM method. The highest-resolution halos considered here, which have $10^6$ particles, are more robust to changes in the initial conditions, but still lead to 10\% level uncertainties in collapse time. This highlights that a certain level of uncertainty persists across our simulations even after attempting to optimize the SIDM implementation.
\end{itemize}

This work underscores the challenges encountered when simulating the SIDM gravothermal collapse process and comparing results across the literature that may start from different sets of simulation parameters, initial conditions and methods.   While we have only focused on an isolated $\sim 10^9 ~M_{\odot}$ halo here as a concrete example, the results already demonstrate that the detailed implementation of the gravitational and self interactions can significantly alter a halo's evolution into the core-collapse regime.  These results motivate further convergence studies for an expanded range of halo masses and should also be generalized to a cosmological setting.  Such work will help minimize numerical mis-modeling effects and ensure that spurious energy gains or losses in SIDM simulations are not misattributed to genuine physical effects.

\vspace{0.1in}
\noindent \textbf{Note Added} \\

\vspace{-0.06in}
\noindent As this paper was being completed, we became aware of Ref.~\cite{Mace:2024uze}, which also studies numerical uncertainties on SIDM halo evolution. Both works consider the convergence of SIDM halos in terms of both mass and time resolution, for which the results are in general agreement. However, our paper  additionally explores the systematic differences that may stem from different SIDM implementations and adaptive force softening.
\acknowledgments
The authors acknowledge Frank~van den Bosch, Akaxia~Cruz, Phil~Hopkins and Annika~Peter for helpful conversations. ML is supported by the Department of Energy~(DOE) under Award Number DE-SC0007968 and the Simons Investigator in Physics Award.  ML and OS were supported by the Binational Science Foundation (grant No. 2018140). OS is also supported by the NSF (grant~No.~PHY-2210498) and acknowledges support from the Yang Institute for Theoretical Physics and the Sarbinowski Foundation. ML and OS acknowledge the Simons Foundation for support. MK is supported by the NSF (grant~No.~PHY-2210283). This work was performed in part at Aspen Center for Physics, which is supported by National Science Foundation grant PHY-2210452. Simulations were performed at the high-performance computing cluster at the Research Cyber
infrastructure Center (RCIC), which provides systems, application software, and scalable storage support to the UCI
research community.

The work presented in this paper was also performed on computational resources managed and supported by Princeton Research Computing. This research made extensive use of the publicly available codes
\texttt{IPython}~\citep{Perez07}, 
\texttt{matplotlib}~\citep{Hunter07}, 
%\texttt{scikit-learn}~\citep{Pedregosa11},
\texttt{Jupyter}~\citep{Kluyver16},
\texttt{NumPy}~\citep{Harris2020},
\texttt{SciPy}~\citep{Virtanen20}, 
and
\texttt{astropy}~\citep{Astropy2022}.


\bibliographystyle{JHEP}
\bibliography{ref.bib} % if your bibtex file is called example.bib

\providecommand{\href}[2]{#2}\begingroup\raggedright\begin{thebibliography}{10}

\bibitem{Spergel2000}
D.N.~{Spergel} and P.J.~{Steinhardt}, \emph{{Observational Evidence for
  Self-Interacting Cold Dark Matter}},
  \href{https://doi.org/10.1103/PhysRevLett.84.3760}{\emph{\prl} {\bfseries 84}
  (2000) 3760} [\href{https://arxiv.org/abs/astro-ph/9909386}{{\ttfamily
  astro-ph/9909386}}].

\bibitem{Tulin:2017ara}
S.~Tulin and H.-B.~Yu, \emph{{Dark Matter Self-interactions and Small Scale
  Structure}}, \href{https://doi.org/10.1016/j.physrep.2017.11.004}{\emph{Phys.
  Rept.} {\bfseries 730} (2018) 1}
  [\href{https://arxiv.org/abs/1705.02358}{{\ttfamily 1705.02358}}].

\bibitem{Adhikari:2022sbh}
S.~Adhikari et~al., \emph{{Astrophysical Tests of Dark Matter
  Self-Interactions}},  \href{https://arxiv.org/abs/2207.10638}{{\ttfamily
  2207.10638}}.

\bibitem{Kochanek2000}
C.S.~{Kochanek} and M.~{White}, \emph{{A Quantitative Study of Interacting Dark
  Matter in Halos}}, \href{https://doi.org/10.1086/317149}{\emph{\apj}
  {\bfseries 543} (2000) 514}
  [\href{https://arxiv.org/abs/astro-ph/0003483}{{\ttfamily
  astro-ph/0003483}}].

\bibitem{Yoshida2000}
N.~{Yoshida}, V.~{Springel}, S.D.M.~{White} and G.~{Tormen}, \emph{{Weakly
  Self-interacting Dark Matter and the Structure of Dark Halos}},
  \href{https://doi.org/10.1086/317306}{\emph{\apjl} {\bfseries 544} (2000)
  L87} [\href{https://arxiv.org/abs/astro-ph/0006134}{{\ttfamily
  astro-ph/0006134}}].

\bibitem{Koda2011}
J.~{Koda} and P.R.~{Shapiro}, \emph{{Gravothermal collapse of isolated
  self-interacting dark matter haloes: N-body simulation versus the fluid
  model}},
  \href{https://doi.org/10.1111/j.1365-2966.2011.18684.x}{\emph{\mnras}
  {\bfseries 415} (2011) 1125}
  [\href{https://arxiv.org/abs/1101.3097}{{\ttfamily 1101.3097}}].

\bibitem{Vogelsberger2012}
M.~{Vogelsberger}, J.~{Zavala} and A.~{Loeb}, \emph{{Subhaloes in
  self-interacting galactic dark matter haloes}},
  \href{https://doi.org/10.1111/j.1365-2966.2012.21182.x}{\emph{\mnras}
  {\bfseries 423} (2012) 3740}
  [\href{https://arxiv.org/abs/1201.5892}{{\ttfamily 1201.5892}}].

\bibitem{Rocha2013}
M.~{Rocha}, A.H.G.~{Peter}, J.S.~{Bullock}, M.~{Kaplinghat},
  S.~{Garrison-Kimmel}, J.~{O{\~n}orbe} et~al., \emph{{Cosmological simulations
  with self-interacting dark matter - I. Constant-density cores and
  substructure}}, \href{https://doi.org/10.1093/mnras/sts514}{\emph{\mnras}
  {\bfseries 430} (2013) 81} [\href{https://arxiv.org/abs/1208.3025}{{\ttfamily
  1208.3025}}].

\bibitem{BalbShap2002}
S.~{Balberg} and S.L.~{Shapiro}, \emph{{Gravothermal Collapse of
  Self-Interacting Dark Matter Halos and the Origin of Massive Black Holes}},
  \href{https://doi.org/10.1103/PhysRevLett.88.101301}{\emph{\prl} {\bfseries
  88} (2002) 101301} [\href{https://arxiv.org/abs/astro-ph/0111176}{{\ttfamily
  astro-ph/0111176}}].

\bibitem{Elbert:2014bma}
O.D.~Elbert, J.S.~Bullock, S.~Garrison-Kimmel, M.~Rocha, J.~O\~norbe and
  A.H.G.~Peter, \emph{{Core formation in dwarf haloes with self-interacting
  dark matter: no fine-tuning necessary}},
  \href{https://doi.org/10.1093/mnras/stv1470}{\emph{Mon. Not. Roy. Astron.
  Soc.} {\bfseries 453} (2015) 29}
  [\href{https://arxiv.org/abs/1412.1477}{{\ttfamily 1412.1477}}].

\bibitem{Lyndel1980}
D.~Lynden-Bell and P.P.~Eggleton, \emph{{On the consequences of the
  gravothermal catastrophe}},
  \href{https://doi.org/10.1093/mnras/191.3.483}{\emph{Monthly Notices of the
  Royal Astronomical Society} {\bfseries 191} (1980) 483}
  [\href{https://arxiv.org/abs/https://academic.oup.com/mnras/article-pdf/191/3/483/2913541/mnras191-0483.pdf}{{\ttfamily
  https://academic.oup.com/mnras/article-pdf/191/3/483/2913541/mnras191-0483.pdf}}].

\bibitem{Essig2019}
R.~{Essig}, S.D.~{McDermott}, H.-B.~{Yu} and Y.-M.~{Zhong}, \emph{{Constraining
  Dissipative Dark Matter Self-Interactions}},
  \href{https://doi.org/10.1103/PhysRevLett.123.121102}{\emph{\prl} {\bfseries
  123} (2019) 121102} [\href{https://arxiv.org/abs/1809.01144}{{\ttfamily
  1809.01144}}].

\bibitem{Nishikawa2020}
H.~{Nishikawa}, K.K.~{Boddy} and M.~{Kaplinghat}, \emph{{Accelerated core
  collapse in tidally stripped self-interacting dark matter halos}},
  \href{https://doi.org/10.1103/PhysRevD.101.063009}{\emph{\prd} {\bfseries
  101} (2020) 063009} [\href{https://arxiv.org/abs/1901.00499}{{\ttfamily
  1901.00499}}].

\bibitem{Yang:2022hkm}
D.~Yang and H.-B.~Yu, \emph{{Gravothermal evolution of dark matter halos with
  differential elastic scattering}},
  \href{https://doi.org/10.1088/1475-7516/2022/09/077}{\emph{JCAP} {\bfseries
  09} (2022) 077} [\href{https://arxiv.org/abs/2205.03392}{{\ttfamily
  2205.03392}}].

\bibitem{Yang2023}
S.~{Yang}, X.~{Du}, Z.C.~{Zeng}, A.~{Benson}, F.~{Jiang}, E.O.~{Nadler} et~al.,
  \emph{{Gravothermal Solutions of SIDM Halos: Mapping from Constant to
  Velocity-dependent Cross Section}},
  \href{https://doi.org/10.3847/1538-4357/acbd49}{\emph{\apj} {\bfseries 946}
  (2023) 47} [\href{https://arxiv.org/abs/2205.02957}{{\ttfamily 2205.02957}}].

\bibitem{Yang:2023jwn}
D.~Yang, E.O.~Nadler, H.-B.~Yu and Y.-M.~Zhong, \emph{{A Parametric Model for
  Self-Interacting Dark Matter Halos}},
  \href{https://arxiv.org/abs/2305.16176}{{\ttfamily 2305.16176}}.

\bibitem{Jiang2023}
F.~{Jiang}, A.~{Benson}, P.F.~{Hopkins}, O.~{Slone}, M.~{Lisanti},
  M.~{Kaplinghat} et~al., \emph{{A semi-analytic study of self-interacting
  dark-matter haloes with baryons}},
  \href{https://doi.org/10.1093/mnras/stad705}{\emph{\mnras} {\bfseries 521}
  (2023) 4630} [\href{https://arxiv.org/abs/2206.12425}{{\ttfamily
  2206.12425}}].

\bibitem{Zhong:2023yzk}
Y.-M.~Zhong, D.~Yang and H.-B.~Yu, \emph{{The impact of baryonic potentials on
  the gravothermal evolution of self-interacting dark matter haloes}},
  \href{https://doi.org/10.1093/mnras/stad2765}{\emph{Mon. Not. Roy. Astron.
  Soc.} {\bfseries 526} (2023) 758}
  [\href{https://arxiv.org/abs/2306.08028}{{\ttfamily 2306.08028}}].

\bibitem{Ren:2018jpt}
T.~Ren, A.~Kwa, M.~Kaplinghat and H.-B.~Yu, \emph{{Reconciling the Diversity
  and Uniformity of Galactic Rotation Curves with Self-Interacting Dark
  Matter}}, \href{https://doi.org/10.1103/PhysRevX.9.031020}{\emph{Phys. Rev.
  X} {\bfseries 9} (2019) 031020}
  [\href{https://arxiv.org/abs/1808.05695}{{\ttfamily 1808.05695}}].

\bibitem{Zentner:2022xux}
A.~Zentner, S.~Dandavate, O.~Slone and M.~Lisanti, \emph{{A critical assessment
  of solutions to the galaxy diversity problem}},
  \href{https://doi.org/10.1088/1475-7516/2022/07/031}{\emph{JCAP} {\bfseries
  07} (2022) 031} [\href{https://arxiv.org/abs/2202.00012}{{\ttfamily
  2202.00012}}].

\bibitem{Kaplinghat2019}
M.~{Kaplinghat}, M.~{Valli} and H.-B.~{Yu}, \emph{{Too big to fail in light of
  Gaia}}, \href{https://doi.org/10.1093/mnras/stz2511}{\emph{\mnras} {\bfseries
  490} (2019) 231} [\href{https://arxiv.org/abs/1904.04939}{{\ttfamily
  1904.04939}}].

\bibitem{Zavala:2019sjk}
J.~Zavala, M.R.~Lovell, M.~Vogelsberger and J.D.~Burger, \emph{{Diverse dark
  matter density at sub-kiloparsec scales in Milky Way satellites: Implications
  for the nature of dark matter}},
  \href{https://doi.org/10.1103/PhysRevD.100.063007}{\emph{Phys. Rev. D}
  {\bfseries 100} (2019) 063007}
  [\href{https://arxiv.org/abs/1904.09998}{{\ttfamily 1904.09998}}].

\bibitem{Kahlhoefer2019}
F.~{Kahlhoefer}, M.~{Kaplinghat}, T.R.~{Slatyer} and C.-L.~{Wu},
  \emph{{Diversity in density profiles of self-interacting dark matter
  satellite halos}},
  \href{https://doi.org/10.1088/1475-7516/2019/12/010}{\emph{\jcap} {\bfseries
  2019} (2019) 010} [\href{https://arxiv.org/abs/1904.10539}{{\ttfamily
  1904.10539}}].

\bibitem{Correa2021}
C.A.~{Correa}, \emph{{Constraining velocity-dependent self-interacting dark
  matter with the Milky Way's dwarf spheroidal galaxies}},
  \href{https://doi.org/10.1093/mnras/stab506}{\emph{\mnras} {\bfseries 503}
  (2021) 920} [\href{https://arxiv.org/abs/2007.02958}{{\ttfamily
  2007.02958}}].

\bibitem{Turner2021}
H.C.~{Turner}, M.R.~{Lovell}, J.~{Zavala} and M.~{Vogelsberger}, \emph{{The
  onset of gravothermal core collapse in velocity-dependent self-interacting
  dark matter subhaloes}},
  \href{https://doi.org/10.1093/mnras/stab1725}{\emph{\mnras} {\bfseries 505}
  (2021) 5327} [\href{https://arxiv.org/abs/2010.02924}{{\ttfamily
  2010.02924}}].

\bibitem{Yang2022}
D.~{Yang}, E.O.~{Nadler} and H.-b.~{Yu}, \emph{{Strong Dark Matter
  Self-interactions Diversify Halo Populations Within and Surrounding the Milky
  Way}}, \href{https://doi.org/10.48550/arXiv.2211.13768}{\emph{arXiv e-prints}
  (2022) arXiv:2211.13768} [\href{https://arxiv.org/abs/2211.13768}{{\ttfamily
  2211.13768}}].

\bibitem{Slone:2021nqd}
O.~Slone, F.~Jiang, M.~Lisanti and M.~Kaplinghat, \emph{{Orbital evolution of
  satellite galaxies in self-interacting dark matter models}},
  \href{https://doi.org/10.1103/PhysRevD.107.043014}{\emph{Phys. Rev. D}
  {\bfseries 107} (2023) 043014}
  [\href{https://arxiv.org/abs/2108.03243}{{\ttfamily 2108.03243}}].

\bibitem{Silverman2022}
M.~Silverman, J.S.~Bullock, M.~Kaplinghat, V.H.~Robles and M.~Valli,
  \emph{Motivations for a large self-interacting dark matter cross-section from
  milky way satellites},
  \href{https://doi.org/10.1093/mnras/stac3232}{\emph{Monthly Notices of the
  Royal Astronomical Society} {\bfseries 518} (2022) 2418–2435}.

\bibitem{Meneghetti2022}
M.~{Meneghetti}, A.~{Ragagnin}, S.~{Borgani}, F.~{Calura}, G.~{Despali},
  C.~{Giocoli} et~al., \emph{{The probability of galaxy-galaxy strong lensing
  events in hydrodynamical simulations of galaxy clusters}},
  \href{https://doi.org/10.1051/0004-6361/202243779}{\emph{\aap} {\bfseries
  668} (2022) A188} [\href{https://arxiv.org/abs/2204.09065}{{\ttfamily
  2204.09065}}].

\bibitem{Yang2021}
D.~{Yang} and H.-B.~{Yu}, \emph{{Self-interacting dark matter and small-scale
  gravitational lenses in galaxy clusters}},
  \href{https://doi.org/10.1103/PhysRevD.104.103031}{\emph{\prd} {\bfseries
  104} (2021) 103031} [\href{https://arxiv.org/abs/2102.02375}{{\ttfamily
  2102.02375}}].

\bibitem{Minor2021}
Q.~{Minor}, S.~{Gad-Nasr}, M.~{Kaplinghat} and S.~{Vegetti}, \emph{{An
  unexpected high concentration for the dark substructure in the gravitational
  lens SDSSJ0946+1006}},
  \href{https://doi.org/10.1093/mnras/stab2247}{\emph{\mnras} {\bfseries 507}
  (2021) 1662} [\href{https://arxiv.org/abs/2011.10627}{{\ttfamily
  2011.10627}}].

\bibitem{Sengul:2022edu}
A.c.~\c{S}eng\"ul and C.~Dvorkin, \emph{{Probing dark matter with strong
  gravitational lensing through an effective density slope}},
  \href{https://doi.org/10.1093/mnras/stac2256}{\emph{Mon. Not. Roy. Astron.
  Soc.} {\bfseries 516} (2022) 336}
  [\href{https://arxiv.org/abs/2206.10635}{{\ttfamily 2206.10635}}].

\bibitem{Zhang:2023wda}
G.~Zhang, A.c.~\c{S}eng\"ul and C.~Dvorkin, \emph{{Subhalo effective density
  slope measurements from HST strong lensing data with neural likelihood-ratio
  estimation}},  \href{https://arxiv.org/abs/2308.09739}{{\ttfamily
  2308.09739}}.

\bibitem{Pollack2015}
J.~{Pollack}, D.N.~{Spergel} and P.J.~{Steinhardt}, \emph{{Supermassive Black
  Holes from Ultra-strongly Self-interacting Dark Matter}},
  \href{https://doi.org/10.1088/0004-637X/804/2/131}{\emph{\apj} {\bfseries
  804} (2015) 131} [\href{https://arxiv.org/abs/1501.00017}{{\ttfamily
  1501.00017}}].

\bibitem{Damico2018}
G.~{D'Amico}, P.~{Panci}, A.~{Lupi}, S.~{Bovino} and J.~{Silk}, \emph{{Massive
  black holes from dissipative dark matter}},
  \href{https://doi.org/10.1093/mnras/stx2419}{\emph{\mnras} {\bfseries 473}
  (2018) 328} [\href{https://arxiv.org/abs/1707.03419}{{\ttfamily
  1707.03419}}].

\bibitem{Latif2019}
M.A.~{Latif}, A.~{Lupi}, D.R.G.~{Schleicher}, G.~{D'Amico}, P.~{Panci} and
  S.~{Bovino}, \emph{{Black hole formation in the context of dissipative dark
  matter}}, \href{https://doi.org/10.1093/mnras/stz608}{\emph{\mnras}
  {\bfseries 485} (2019) 3352}
  [\href{https://arxiv.org/abs/1812.03104}{{\ttfamily 1812.03104}}].

\bibitem{Choquette2019}
J.~{Choquette}, J.M.~{Cline} and J.M.~{Cornell}, \emph{{Early formation of
  supermassive black holes via dark matter self-interactions}},
  \href{https://doi.org/10.1088/1475-7516/2019/07/036}{\emph{\jcap} {\bfseries
  2019} (2019) 036} [\href{https://arxiv.org/abs/1812.05088}{{\ttfamily
  1812.05088}}].

\bibitem{Feng2021}
W.-X.~{Feng}, H.-B.~{Yu} and Y.-M.~{Zhong}, \emph{{Seeding Supermassive Black
  Holes with Self-interacting Dark Matter: A Unified Scenario with Baryons}},
  \href{https://doi.org/10.3847/2041-8213/ac04b0}{\emph{\apjl} {\bfseries 914}
  (2021) L26} [\href{https://arxiv.org/abs/2010.15132}{{\ttfamily
  2010.15132}}].

\bibitem{Blinnikov1983}
S.I.~{Blinnikov} and M.Y.~{Khlopov}, \emph{{Possible Astronomical Effects of
  Mirror Particles}}, {\emph{\sovast} {\bfseries 27} (1983) 371}.

\bibitem{Klhopov1991}
M.Y.~{Khlopov}, G.M.~{Beskin}, N.G.~{Bochkarev}, L.A.~{Pustilnik} and
  S.A.~{Pustilnik}, \emph{{Observational Physics of the Mirror World}},
  {\emph{\sovast} {\bfseries 35} (1991) 21}.

\bibitem{Foot2004}
R.~{Foot}, \emph{{Mirror Matter-Type Dark Matter}},
  \href{https://doi.org/10.1142/S0218271804006449}{\emph{International Journal
  of Modern Physics D} {\bfseries 13} (2004) 2161}
  [\href{https://arxiv.org/abs/astro-ph/0407623}{{\ttfamily
  astro-ph/0407623}}].

\bibitem{Berezhiani2005}
Z.~{Berezhiani}, \emph{{Through the Looking-Glass Alice's Adventures in Mirror
  World}},  in \emph{From Fields to Strings: Circumnavigating Theoretical
  Physics: Ian Kogan Memorial Collection (in 3 Vols). Edited by SHIFMAN MISHA
  ET AL. Published by World Scientific Publishing Co. Pte. Ltd}, pp.~2147--2195
  (2005), \href{https://doi.org/10.1142/9789812775344_0055}{DOI}.

\bibitem{Fry2015}
A.B.~{Fry}, F.~{Governato}, A.~{Pontzen}, T.~{Quinn}, M.~{Tremmel},
  L.~{Anderson} et~al., \emph{{All about baryons: revisiting SIDM predictions
  at small halo masses}},
  \href{https://doi.org/10.1093/mnras/stv1330}{\emph{\mnras} {\bfseries 452}
  (2015) 1468} [\href{https://arxiv.org/abs/1501.00497}{{\ttfamily
  1501.00497}}].

\bibitem{Robles2017}
V.H.~{Robles}, J.S.~{Bullock}, O.D.~{Elbert}, A.~{Fitts},
  A.~{Gonz{\'a}lez-Samaniego}, M.~{Boylan-Kolchin} et~al., \emph{{SIDM on FIRE:
  hydrodynamical self-interacting dark matter simulations of low-mass dwarf
  galaxies}}, \href{https://doi.org/10.1093/mnras/stx2253}{\emph{\mnras}
  {\bfseries 472} (2017) 2945}
  [\href{https://arxiv.org/abs/1706.07514}{{\ttfamily 1706.07514}}].

\bibitem{Robertson2017a}
A.~{Robertson}, R.~{Massey} and V.~{Eke}, \emph{{What does the Bullet Cluster
  tell us about self-interacting dark matter?}},
  \href{https://doi.org/10.1093/mnras/stw2670}{\emph{\mnras} {\bfseries 465}
  (2017) 569} [\href{https://arxiv.org/abs/1605.04307}{{\ttfamily
  1605.04307}}].

\bibitem{Robertson2017b}
A.~{Robertson}, R.~{Massey} and V.~{Eke}, \emph{{Cosmic particle colliders:
  simulations of self-interacting dark matter with anisotropic scattering}},
  \href{https://doi.org/10.1093/mnras/stx463}{\emph{\mnras} {\bfseries 467}
  (2017) 4719} [\href{https://arxiv.org/abs/1612.03906}{{\ttfamily
  1612.03906}}].

\bibitem{Nadler2020}
E.O.~{Nadler}, A.~{Banerjee}, S.~{Adhikari}, Y.-Y.~{Mao} and R.H.~{Wechsler},
  \emph{{Signatures of Velocity-dependent Dark Matter Self-interactions in
  Milky Way-mass Halos}},
  \href{https://doi.org/10.3847/1538-4357/ab94b0}{\emph{\apj} {\bfseries 896}
  (2020) 112} [\href{https://arxiv.org/abs/2001.08754}{{\ttfamily
  2001.08754}}].

\bibitem{Banerjee2020}
A.~{Banerjee}, S.~{Adhikari}, N.~{Dalal}, S.~{More} and A.~{Kravtsov},
  \emph{{Signatures of self-interacting dark matter on cluster density profile
  and subhalo distributions}},
  \href{https://doi.org/10.1088/1475-7516/2020/02/024}{\emph{\jcap} {\bfseries
  2020} (2020) 024} [\href{https://arxiv.org/abs/1906.12026}{{\ttfamily
  1906.12026}}].

\bibitem{Correa2022}
C.A.~{Correa}, M.~{Schaller}, S.~{Ploeckinger}, N.~{Anau Montel}, C.~{Weniger}
  and S.~{Ando}, \emph{{TangoSIDM: tantalizing models of self-interacting dark
  matter}}, \href{https://doi.org/10.1093/mnras/stac2830}{\emph{\mnras}
  {\bfseries 517} (2022) 3045}
  [\href{https://arxiv.org/abs/2206.11298}{{\ttfamily 2206.11298}}].

\bibitem{Hernquist1990}
L.~{Hernquist}, \emph{{An Analytical Model for Spherical Galaxies and Bulges}},
  \href{https://doi.org/10.1086/168845}{\emph{\apj} {\bfseries 356} (1990)
  359}.

\bibitem{Navarro1997}
J.F.~{Navarro}, C.S.~{Frenk} and S.D.M.~{White}, \emph{{A Universal Density
  Profile from Hierarchical Clustering}},
  \href{https://doi.org/10.1086/304888}{\emph{\apj} {\bfseries 490} (1997) 493}
  [\href{https://arxiv.org/abs/astro-ph/9611107}{{\ttfamily
  astro-ph/9611107}}].

\bibitem{Meskhidze2022}
H.~{Meskhidze}, F.J.~{Mercado}, O.~{Sameie}, V.H.~{Robles}, J.S.~{Bullock},
  M.~{Kaplinghat} et~al., \emph{{Comparing implementations of self-interacting
  dark matter in the GIZMO and AREPO codes}},
  \href{https://doi.org/10.1093/mnras/stac1056}{\emph{\mnras} {\bfseries 513}
  (2022) 2600} [\href{https://arxiv.org/abs/2203.06035}{{\ttfamily
  2203.06035}}].

\bibitem{Hopkins2015}
P.F.~{Hopkins}, \emph{{A new class of accurate, mesh-free hydrodynamic
  simulation methods}},
  \href{https://doi.org/10.1093/mnras/stv195}{\emph{\mnras} {\bfseries 450}
  (2015) 53} [\href{https://arxiv.org/abs/1409.7395}{{\ttfamily 1409.7395}}].

\bibitem{Garrison2013}
S.~{Garrison-Kimmel}, M.~{Rocha}, M.~{Boylan-Kolchin}, J.S.~{Bullock} and
  J.~{Lally}, \emph{{Can feedback solve the too-big-to-fail problem?}},
  \href{https://doi.org/10.1093/mnras/stt984}{\emph{\mnras} {\bfseries 433}
  (2013) 3539} [\href{https://arxiv.org/abs/1301.3137}{{\ttfamily 1301.3137}}].

\bibitem{Zemp2008}
M.~{Zemp}, B.~{Moore}, J.~{Stadel}, C.M.~{Carollo} and P.~{Madau},
  \emph{{Multimass spherical structure models for N-body simulations}},
  \href{https://doi.org/10.1111/j.1365-2966.2008.13126.x}{\emph{\mnras}
  {\bfseries 386} (2008) 1543}
  [\href{https://arxiv.org/abs/0710.3189}{{\ttfamily 0710.3189}}].

\bibitem{Zeng2022}
Z.C.~{Zeng}, A.H.G.~{Peter}, X.~{Du}, A.~{Benson}, S.~{Kim}, F.~{Jiang} et~al.,
  \emph{{Core-collapse, evaporation, and tidal effects: the life story of a
  self-interacting dark matter subhalo}},
  \href{https://doi.org/10.1093/mnras/stac1094}{\emph{\mnras} {\bfseries 513}
  (2022) 4845} [\href{https://arxiv.org/abs/2110.00259}{{\ttfamily
  2110.00259}}].

\bibitem{Springel2005}
V.~{Springel}, \emph{{The cosmological simulation code GADGET-2}},
  \href{https://doi.org/10.1111/j.1365-2966.2005.09655.x}{\emph{\mnras}
  {\bfseries 364} (2005) 1105}
  [\href{https://arxiv.org/abs/astro-ph/0505010}{{\ttfamily
  astro-ph/0505010}}].

\bibitem{Hopkins2018}
P.F.~{Hopkins}, A.~{Wetzel}, D.~{Kere{\v{s}}}, C.-A.~{Faucher-Gigu{\`e}re},
  E.~{Quataert}, M.~{Boylan-Kolchin} et~al., \emph{{FIRE-2 simulations: physics
  versus numerics in galaxy formation}},
  \href{https://doi.org/10.1093/mnras/sty1690}{\emph{\mnras} {\bfseries 480}
  (2018) 800} [\href{https://arxiv.org/abs/1702.06148}{{\ttfamily
  1702.06148}}].

\bibitem{vdBosch2018}
F.C.~{van den Bosch} and G.~{Ogiya}, \emph{{Dark matter substructure in
  numerical simulations: a tale of discreteness noise, runaway instabilities,
  and artificial disruption}},
  \href{https://doi.org/10.1093/mnras/sty084}{\emph{\mnras} {\bfseries 475}
  (2018) 4066} [\href{https://arxiv.org/abs/1801.05427}{{\ttfamily
  1801.05427}}].

\bibitem{Power2003}
C.~{Power}, J.F.~{Navarro}, A.~{Jenkins}, C.S.~{Frenk}, S.D.M.~{White},
  V.~{Springel} et~al., \emph{{The inner structure of {\ensuremath{\Lambda}}CDM
  haloes - I. A numerical convergence study}},
  \href{https://doi.org/10.1046/j.1365-8711.2003.05925.x}{\emph{\mnras}
  {\bfseries 338} (2003) 14}
  [\href{https://arxiv.org/abs/astro-ph/0201544}{{\ttfamily
  astro-ph/0201544}}].

\bibitem{Balberg2002}
S.~{Balberg}, S.L.~{Shapiro} and S.~{Inagaki}, \emph{{Self-Interacting Dark
  Matter Halos and the Gravothermal Catastrophe}},
  \href{https://doi.org/10.1086/339038}{\emph{\apj} {\bfseries 568} (2002) 475}
  [\href{https://arxiv.org/abs/astro-ph/0110561}{{\ttfamily
  astro-ph/0110561}}].

\bibitem{Outmezguine2022}
N.J.~{Outmezguine}, K.K.~{Boddy}, S.~{Gad-Nasr}, M.~{Kaplinghat} and
  L.~{Sagunski}, \emph{{Universal gravothermal evolution of isolated
  self-interacting dark matter halos for velocity-dependent cross sections}},
  {\emph{arXiv e-prints} (2022) arXiv:2204.06568}
  [\href{https://arxiv.org/abs/2204.06568}{{\ttfamily 2204.06568}}].

\bibitem{Fischer2021}
M.S.~{Fischer}, M.~{Br{\"u}ggen}, K.~{Schmidt-Hoberg}, K.~{Dolag},
  F.~{Kahlhoefer}, A.~{Ragagnin} et~al., \emph{{N-body simulations of dark
  matter with frequent self-interactions}},
  \href{https://doi.org/10.1093/mnras/stab1198}{\emph{\mnras} {\bfseries 505}
  (2021) 851} [\href{https://arxiv.org/abs/2012.10277}{{\ttfamily
  2012.10277}}].

\bibitem{Elbert2015}
O.D.~{Elbert}, J.S.~{Bullock}, S.~{Garrison-Kimmel}, M.~{Rocha},
  J.~{O{\~n}orbe} and A.H.G.~{Peter}, \emph{{Core formation in dwarf haloes
  with self-interacting dark matter: no fine-tuning necessary}},
  \href{https://doi.org/10.1093/mnras/stv1470}{\emph{\mnras} {\bfseries 453}
  (2015) 29} [\href{https://arxiv.org/abs/1412.1477}{{\ttfamily 1412.1477}}].

\bibitem{Mace:2024uze}
C.~Mace, Z.C.~Zeng, A.H.G.~Peter, X.~Du, S.~Yang, A.~Benson et~al.,
  \emph{{Convergence Tests of Self-Interacting Dark Matter Simulations}},
  \href{https://arxiv.org/abs/2402.01604}{{\ttfamily 2402.01604}}.

\bibitem{Perez07}
F.~{Perez} and B.E.~{Granger}, \emph{{IPython: A System for Interactive
  Scientific Computing}},
  \href{https://doi.org/10.1109/MCSE.2007.53}{\emph{Computing in Science and
  Engineering} {\bfseries 9} (2007) 21}.

\bibitem{Hunter07}
J.D.~{Hunter}, \emph{{Matplotlib: A 2D Graphics Environment}},
  \href{https://doi.org/10.1109/MCSE.2007.55}{\emph{Computing in Science and
  Engineering} {\bfseries 9} (2007) 90}.

\bibitem{Kluyver16}
T.~{Kluyver}, B.~{Ragan-Kelley}, F.~{P{\'e}rez}, B.~{Granger}, M.~{Bussonnier},
  J.~{Frederic} et~al., \emph{{Jupyter Notebooks{\textemdash}a publishing
  format for reproducible computational workflows}},  in \emph{IOS Press},
  pp.~87--90 (2016), \href{https://doi.org/10.3233/978-1-61499-649-1-87}{DOI}.

\bibitem{Harris2020}
C.R.~{Harris}, K.J.~{Millman}, S.J.~{van der Walt}, R.~{Gommers},
  P.~{Virtanen}, D.~{Cournapeau} et~al., \emph{{Array programming with NumPy}},
  \href{https://doi.org/10.1038/s41586-020-2649-2}{\emph{\nat} {\bfseries 585}
  (2020) 357} [\href{https://arxiv.org/abs/2006.10256}{{\ttfamily
  2006.10256}}].

\bibitem{Virtanen20}
P.~{Virtanen}, R.~{Gommers}, T.E.~{Oliphant}, M.~{Haberland}, T.~{Reddy},
  D.~{Cournapeau} et~al., \emph{{SciPy 1.0: fundamental algorithms for
  scientific computing in Python}},
  \href{https://doi.org/10.1038/s41592-019-0686-2}{\emph{Nature Methods}
  {\bfseries 17} (2020) 261}
  [\href{https://arxiv.org/abs/1907.10121}{{\ttfamily 1907.10121}}].

\bibitem{Astropy2022}
{Astropy Collaboration}, A.M.~{Price-Whelan}, P.L.~{Lim}, N.~{Earl},
  N.~{Starkman}, L.~{Bradley} et~al., \emph{{The Astropy Project: Sustaining
  and Growing a Community-oriented Open-source Project and the Latest Major
  Release (v5.0) of the Core Package}},
  \href{https://doi.org/10.3847/1538-4357/ac7c74}{\emph{\apj} {\bfseries 935}
  (2022) 167} [\href{https://arxiv.org/abs/2206.14220}{{\ttfamily
  2206.14220}}].

\end{thebibliography}\endgroup

%%%%%%%%%%%%%%%%%%%%%%%%%%%%%%%%%%%%%%%%%%%%%%%%%%

%%%%%%%%%%%%%%%%% APPENDICES %%%%%%%%%%%%%%%%%%%%%
\appendix
\onecolumn
\clearpage
\section{Supplementary Figures}\label{appendix}

\setcounter{equation}{0}
\setcounter{figure}{0} 
\setcounter{table}{0}
\renewcommand{\theequation}{A\arabic{equation}}
\renewcommand{\thefigure}{A\arabic{figure}}
\renewcommand{\thetable}{A\arabic{table}}

\begin{figure}[ht]
    \centering
    \includegraphics[width=\textwidth]{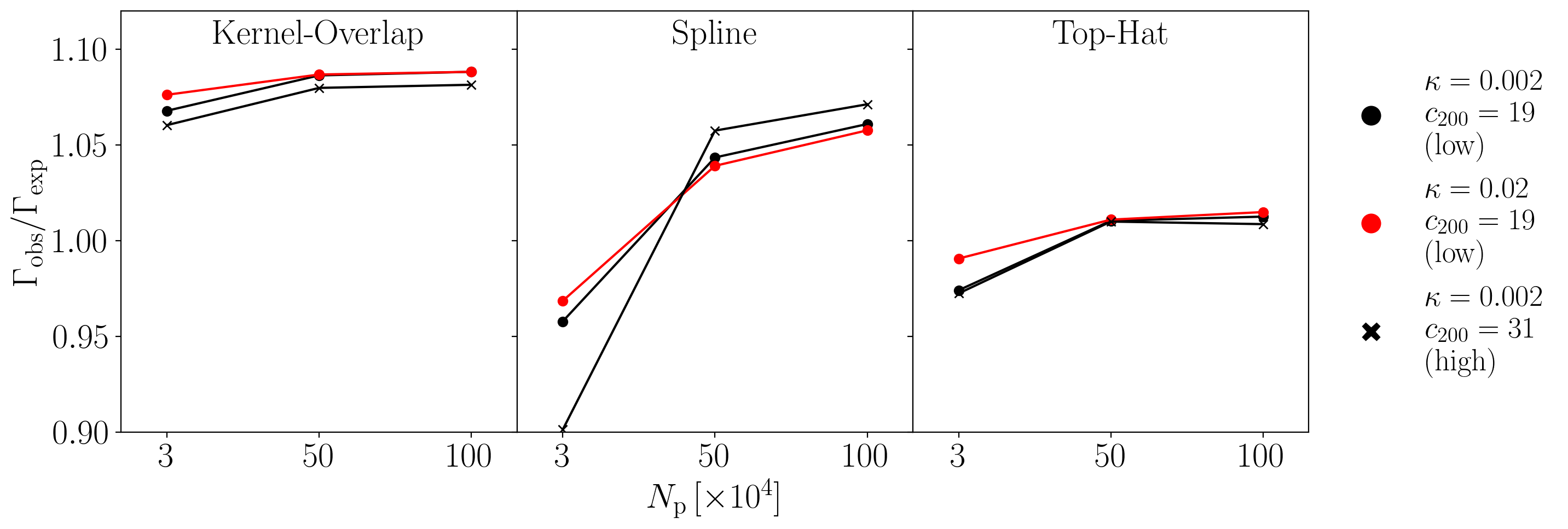}
    \caption{The observed scattering rate, $\Gamma_{\rm obs}$, as a fraction of the expected rate, $\Gamma_{\rm exp}$, plotted as a function of number of particles, $N_p$.  The left, middle and right columns correspond to the Kernel-Overlap, Spline, and Top-Hat results, respectively. Results for the low-concentration halo with $\kappa=0.002$ and $0.02$ are shown by the filled black and red circles, respectively. Results for the high-concentration $\kappa=0.002$ halo are shown by the black crosses.  In general, $\Gamma_{\rm obs}/\Gamma_{\rm exp}$ approaches a common value as the resolution improves.  At the highest resolution simulated here, there are still systematic offsets between the observed and expected scattering rate, with the magnitude depending on the specific SIDM implementation.}
    \label{fig:r/r_e}
\end{figure}

\begin{figure}[ht]
    \centering
    \includegraphics[width=\textwidth]{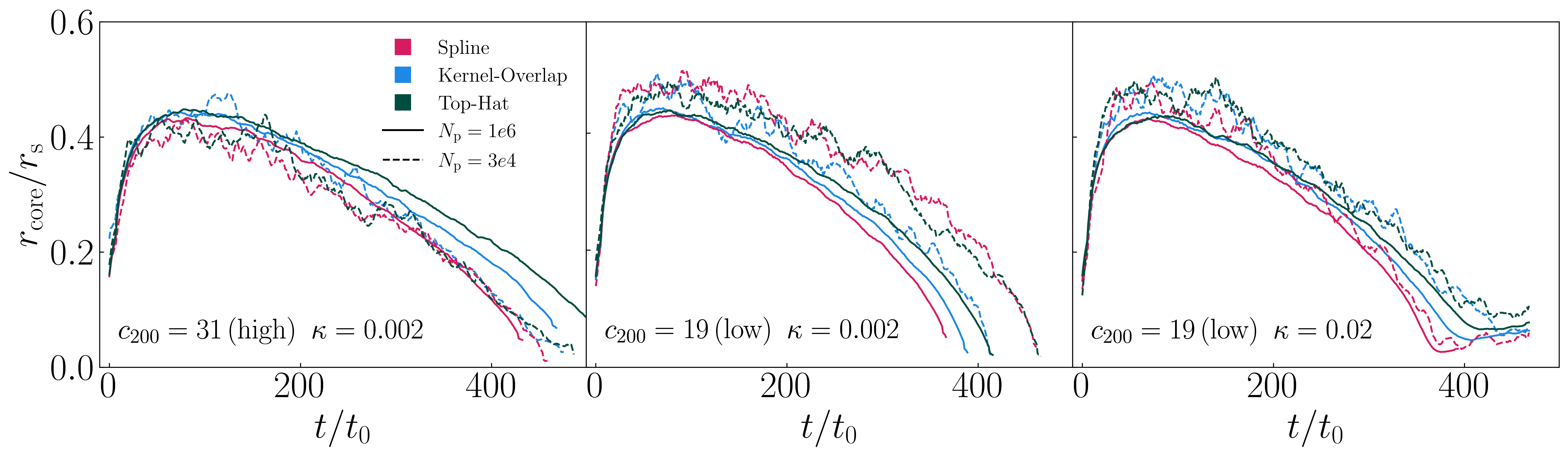}
    \caption{The evolution of the halo's core size, $r_{\rm core}$, rescaled to its scale radius, $r_{\rm s}$.  Results are shown for the low- and high-concentration halos and for different time-stepping criteria.  Solid~(dashed) lines correspond to the highest~(lowest)-resolution simulations run here. The Spline, Kernel-Overlap, and Top-Hat results are shown by the red, blue, and green lines, respectively. The simulations with $\kappa = 0.02$~(right panel) show a visible stalling of the core collapse as the core size grows more slowly at late times in the halo's evolution.}
    \label{fig:coresize}
\end{figure}

\end{document}